\documentclass [11pt]{article}

\ifx\Red\relax\def\Red#1{#1}\fi
\def\Red#1{#1}

\usepackage{epsfig}
\usepackage{amssymb}
\usepackage{verbatim}
\usepackage{delarray}
\usepackage{graphics}
\usepackage{epic}

\newcommand{\lra}{\leftrightarrow}
\newcommand{\inner}[2]{\langle{#1},{#2}\rangle}

\newcommand{\A}{{\mathcal{A}}}

\newcommand{\CC}{{\mathcal{C}}}
\newcommand{\D}{{\mathcal{D}}}
\newcommand{\E}{{\mathcal{E}}}
\newcommand{\FF}{{\mathcal{F}}}

\newcommand{\I}{{\mathcal{I}}}

\newcommand{\SSS}{{\mathcal{S}}}

\newcommand{\C}{{\mathbb{C}}}
\newcommand{\F}{{\mathbb{F}}}

\newcommand{\Z}{{\mathbb{Z}}}
\newcommand{\zerob}{{\mathbf 0}}
\newcommand{\ab}{{\mathbf a}}
\newcommand{\bb}{{\mathbf b}}

\newcommand{\fb}{{\mathbf f}}

\renewcommand{\sb}{{\mathbf s}}

\newcommand{\xb}{{\mathbf x}}

\newcommand{\Ab}{{\mathbf A}}

\newcommand{\Fb}{{\mathbf F}}

\newcommand{\Sb}{{\mathbf S}}

\newcommand{\Xb}{{\mathbf X}}

\newcommand{\Bf}{{\mathfrak{B}}}

\newcommand{\ie}{{\em i.e., }}
\newcommand{\eg}{{\em e.g., }}
\newcommand{\cf}{\emph{cf.\ }}
\newcommand{\etal}{\emph{et al.\ }}

\newcommand{\Tr}{\mathrm{Tr~}}

\newcommand{\half}{\frac{1}{2}}

\newcommand{\dfree}{d_{\mathrm{free}}}

\newcommand{\openbox}{\leavevmode
     \hbox to.77778em{%
     \hfil\vrule
     \vbox to.675em{\hrule width.6em\vfil\hrule}%
     \vrule\hfil}}
\newcommand{\qed}{\hspace*{1cm}\hspace*{\fill}\openbox}

\begin{document}

\title{Codes on Graphs:  Duality and MacWilliams Identities}

\author{G. David Forney, Jr.\footnote{The author is with the
Laboratory for Information and Decision Systems,
Massachusetts Institute of Technology,
Cambridge, MA 02139 (email: forneyd@comcast.net).}}
\date{}

\maketitle

\vspace{-2ex}
\begin{abstract}
A conceptual  framework involving partition functions of normal factor graphs is introduced, paralleling a similar recent development by Al-Bashabsheh and Mao.  The partition functions of dual normal factor graphs are shown to be a Fourier transform pair, whether or not the graphs have cycles.  The original normal graph duality theorem follows as a corollary.

Within this framework, MacWilliams identities are found for various local and global weight generating functions of general group or linear codes on graphs;  this generalizes and provides a concise proof of the MacWilliams identity for linear time-invariant convolutional codes that was recently found by Gluesing-Luerssen and Schneider.  Further MacWilliams identities are developed for terminated convolutional codes, particularly for tail-biting codes, similar to those studied recently by Bocharova, Hug, Johannesson and Kudryashov.
\end{abstract}


\section{Introduction}

Linear codes defined by graphical models have become the central subject of modern coding theory.  Moreover, the topic of ``codes on graphs" has proved to have rich connections to such cognate fields as system theory and statistical physics.

Duality has long been a prominent theme in the study of linear codes.  Duality theory often yields simple, powerful, and surprisingly general results.  Some problems become much simpler in the dual domain.

A fundamental duality result in the field of ``codes on graphs" is the normal graph duality theorem of \cite{F01}.  By imposing  certain natural ``normal" degree restrictions on graphical models, which are in fact not at all restrictive, the normal graph duality theorem shows how a graphical model for a dual linear code $\CC^\perp$ may be obtained by local dualization of a graphical model for a linear code $\CC$, whether or not the graph has cycles. 

At the same time as \cite{F01} (in an adjacent paper in the same special issue), the conceptual framework of ``factor graphs" was introduced by Kschischang, Frey, and Loeliger  \cite{KFL01} to unify various styles of graphical models such as Tanner graphs, Bayesian networks, Markov random fields, Kalman filtering, and so forth, and the various computational algorithms that have been developed independently in these various fields.
Subsequently, the ``normal" restriction has been adopted to some extent in the factor graph literature, particularly by Loeliger and his co-authors (see, \eg \cite{L04, L07}).

Mao and Kschischang \cite{MK05} have developed a general duality theory for factor graphs, particularly by introducing ``convolutional" factor graphs as duals to conventional multiplicative factor graphs.  By specializing to normal graphs, they proved the normal graph duality theorem as a corollary.

Very recently, Al-Bashabsheh and Mao \cite{AM10} have shown that the normal graph duality theorem and Valiant's ``holographic" algorithms can both be united within a common framework, which they call ``normal factor graphs and holographic transformations."  We have collaborated intensively with the authors of that paper and with the Associate Editor of both (P. Vontobel) while writing this paper, and have tried to achieve as much commonality as possible in our presentations.

%\pagebreak

As in \cite{AM10}, we regard normal factor graphs as representations of realizations that involve external variables, internal variables and factors.  All variables are vector spaces over a given finite field $\F$, and all factors are complex-valued.  Each factor involves some subset of the variables, with the following ``normal" degree restrictions:  each external variable is involved in precisely one factor, and each internal variable is involved in precisely two factors.  As shown in \cite{F01}, \emph{any} realization may be straightforwardly ``normalized" by a simple replication procedure that does not affect the topology of the associated graphical model.  In the corresponding normal factor graph (NFG), factors are then represented by vertices, internal variables by ordinary edges of degree 2, and external variables by edges of degree 1 (``half-edges," ``dangling edges," ``dongles").

We define the \emph{partition function} of such a normal factor graph as the sum over all internal variables of the product of all factors.  (In \cite{AM10}, this is called the ``exterior function" of the NFG.)  In particular, if all factors are indicator functions of local linear constraint codes over $\F$, then the partition function is (up to scale) the indicator function of a global linear code $\CC$.

With \cite{AM10}, we believe that the conceptual framework of representing sums of products as partition functions of normal factor graphs is an important paradigm that, surprisingly, does not seem to have been discussed very explicitly in the previous factor graph literature.  We therefore present this framework in some generality in Sections 2 and 3.

In Section 3, we prove a general normal factor graph duality theorem (also proved in \cite{AM10}), which shows that the partition functions of a normal factor graph and its dual are a Fourier transform pair, up to scale.  Specializing to indicator functions of linear codes, we obtain as a corollary a result that is equivalent to the normal graph duality theorem of \cite{F01}.

In Sections 4 and 5, we present further applications that were stimulated by several recent results on MacWilliams identities for linear convolutional codes \cite{GLS08, GLS09, BHJK}.  

For linear block codes, MacWilliams identities are classical duality results that relate weight generating functions of linear codes and their duals $\cite{MS77}$.  
It was shown more than thirty years ago by Shearer and McEliece \cite{SM77} that there is no MacWilliams identity for the usual weight generating function (the free distance spectrum) of a convolutional code.  

However, Gluesing-Luerssen and Schneider (GLS) have recently formulated \cite{GLS08} and proved  \cite{GLS09} a MacWilliams identity that involves the Hamming weight adjacency matrix (HWAM) of a linear time-invariant convolutional code over a finite field and the HWAM of its dual code.  

In Section 2, we show how weight generating functions of various types may be naturally represented as partition functions of normal factor graphs.  Furthermore, for a convolutional code, it is natural to replace a weight generating function (WGF) by a weight adjacency matrix (WAM).  

In Section 4, we then apply our normal factor graph duality results to prove an appropriate MacWilliams identity between a WGF (or WAM) of a local linear code and a WGF (or WAM) of its dual.  We consider exact, complete, and Hamming weight generating functions. This gives a concise proof of the GLS result, and generalizes it to arbitrary group codes defined on graphs;  \eg linear block codes defined on trellises, linear tail-biting codes, or trellis codes over abelian groups.
A preliminary version of these results was presented in \cite{F09}.

\pagebreak 
Bocharova, Hug, Johannesson, and Kudryashov \cite{BHJK} have recently proved a MacWilliams identity that holds for truncations of a convolutional code $\CC$ and its orthogonal code $\CC^\perp$.  As the truncation length becomes large, they obtain approximations to the free distance spectra of $\CC$ and $\CC^\perp$.

In Section 5, we develop MacWilliams identities for distance distributions of various kinds of terminated convolutional codes, particularly tail-biting terminated codes.  A preliminary version of these results was presented in \cite{F10}. These results effectively answer the original question posed by Shearer and McEliece \cite{SM77}. 

%In an appendix, we use a similar argument to provide a concise and transparent proof of the dual sum-product update rule stated in \cite{F01};  see also \cite{BW02}.

\section{Codes, Realizations and Graphical Models}

In this section we review linear codes, realizations of codes, and their graphical models. We start with the development and notation of \cite{F01}, but then transition to normal factor graphs rather than normal graphs.  We develop the general framework of partition functions of normal factor graphs.  Finally, we show how weight generating functions are naturally represented in this framework.

\subsection{Linear codes, realizations and normal graphs}

In this paper we will be concerned with linear codes over a finite field $\F$.  Everything generalizes to group codes over finite abelian groups, but for simplicity we will restrict attention to linear codes.

A linear code $\CC$ over $\F$ is defined over a discrete index set $\I_\A$ and a set $\{A_k: k \in \I_\A\}$ of \emph{symbol alphabets} $A_k$, each a finite-dimensional vector space over $\F$, and thus finite.  The code $\CC$ is then a subspace of the Cartesian-product vector space $\A = \Pi_{k \in \I} A_k$, called the \emph{symbol sequence space}.  

In general, the index set $\I_\A$ may be infinite;  however, in this paper we will assume for simplicity that $\I_\A$ is finite, so that the symbol sequence space $\A$ is finite.  For convolutional codes, this assumption may be justified by considering an infinitely long convolutional code as a limit of a sequence of finitely long terminated codes;  see Section 5.

A \emph{realization} of a code $\CC$ is a concrete implementation or characterization of it.  For instance, a parity-check realization of a linear $(n,k)$ code $\CC$ characterizes it as the set of all $\ab \in \F^n$ that satisfy a set of $n-k$ parity-check equations.  A \emph{Tanner graph} is a graphical model of such a realization.

More generally, a \emph{behavioral realization} of $\CC$ involves not only the set $\{A_k: k \in \I_\A\}$ of symbol alphabets, but also a set $\{S_j: j \in \I_\SSS\}$ of auxiliary alphabets, often called \emph{state spaces}, indexed by a state index set $\I_\SSS$, and a set $\{\CC_i: i \in \I_\CC\}$  of local constraint codes $\CC_i$ indexed by a constraint index set $\I_\CC$, where each constraint code $\CC_i$ involves some subsets $\ab_i$ and $\sb_i$ of the symbol and state variables, respectively.  In a linear behavioral realization, each state space $S_j$ and each constraint code $\CC_i$ is a vector space over $\F$.  We define the state sequence space as $\SSS = \Pi_{j \in \I_\SSS} S_j$.   

The \emph{full behavior} of the realization is the set $\Bf$ of all pairs  $(\ab, \sb) \in \A \times \SSS$ such that all constraints are satisfied;  \ie $(\ab_i, \sb_i) \in \CC_i, \forall i \in \I_\CC$.  The \emph{code} $\CC$ generated by the realization is then the set of all symbol sequences $\ab \in \A$ that appear in some $(\ab, \sb) \in \Bf$.

For example, in a \emph{conventional state realization} of a linear code $\CC$, the symbol index set $\I_\A$ is a conventional  discrete time axis, namely the set of integers $\Z$, or a subinterval of $\Z$.  The state index set $\I_\SSS$ may be thought of as the set of times that occur \emph{between} consecutive pairs of times in $\I_\A$, and the state time preceding symbol time $k \in \I_\A$ is conventionally also denoted by $k \in \I_\SSS$.  The constraint codes $\{\CC_k: k \in I_\A\}$ are linear codes indexed by the symbol index set $\I_\A$, and specify the set of all valid $(s_k, a_k, s_{k+1})$ transitions;  \ie for each $k \in \I_\A$, $\CC_k$ is a subspace of the vector space $S_k \times A_k \times S_{k+1}$.
The full behavior $\Bf$ of the realization is the set of all symbol/state trajectories $(\ab, \sb)$ such that $(s_k, a_k, s_{k+1})$ is a valid transition in $\CC_k$ for all $k \in \I_\A$.  The code $\CC$ generated by the realization is the set of all symbol trajectories $\ab$ that appear  in some $(\ab, \sb) \in \Bf$.

%\pagebreak
A \emph{normal behavioral realization} is defined as a behavioral realization in which every symbol alphabet is involved  in precisely one constraint code, and every state space is involved in precisely two constraint codes.  Thus a conventional state realization is normal.  As shown in \cite{F01}, any behavioral realization may be straightforwardly converted to a normal realization by the ``normalization" procedure that will be described in the next subsection, without essentially increasing the complexity of the realization.

%\pagebreak

%\subsection{Graphical models}

\begin{figure}[h]
\setlength{\unitlength}{5pt}
\centering
\begin{picture}(50,8)(-2, 4)
\multiput(0,5)(12,0){4}{\line(1,0){7}}
\multiput(9.5,7.5)(12,0){3}{\line(0,1){3}}
\multiput(8,10.5)(12,0){3}{\line(1,0){3}}
\put(-3,5){\ldots}
\put(3,6){$S_k$}
\put(13,6){$S_{k+1}$}
\put(25,6){$S_{k+2}$}
\put(37,6){$S_{k+3}$}
\put(44,5){\ldots}
\put(8,11.5){$A_k$}
\put(19,11.5){$A_{k+1}$}
\put(31,11.5){$A_{k+2}$}
\put(7,2.5){\framebox(5,5){$\CC_k$}}
\put(19,2.5){\framebox(5,5){$\CC_{k+1}$}}
\put(31,2.5){\framebox(5,5){$\CC_{k+2}$}}
\end{picture}
\caption{Normal graph of a conventional state realization.}
\label{Fig1}
\end{figure}
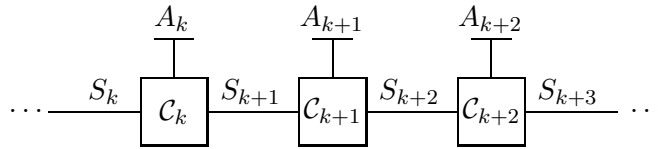

A normal behavioral realization has a natural graphical model, called a \emph{normal graph}, in which each constraint code $\CC_i$ corresponds to a vertex, each state space $S_j$ (which by definition is involved in two constraints) corresponds to an edge connecting the two corresponding constraint vertices, and each symbol alphabet $A_k$  (which by definition is involved in one constraint) corresponds to a leaf or ``half-edge" connected to the corresponding constraint vertex. 
%
% \pagebreak
For example, Figure \ref{Fig1} shows the normal graph corresponding to a conventional state realization, which is a simple chain graph.  Here vertices are represented by square boxes, and the ``half-edges" corresponding to symbol alphabets are represented by special ``dongle" symbols. 

%\pagebreak

%Commonly a linear code $\CC$ is represented by a \emph{Tanner graph}, which is a bipartite graph in which one set of nodes represents the code variables, and another represents local constraints, which are often simple parity checks.  For example, Figure 1(a) shows a Tanner graph for the $(8, 4)$ binary linear extended Hamming code, which has eight variable nodes and four check nodes.  The code $\CC$ consists of all assignments to the variable nodes such that all local constraints are satisfied.

%Alternatively, a code $\CC$ may be represented by a \emph{normal graph}, which is a graph consisting of nodes, ordinary edges connecting pairs of nodes, and half-edges that are incident on only one node.  The ordinary edges represent internal variables, the half-edges represent external variables, and the nodes represent local constraints.  For example, Figure 1(b) shows a normal graph that realizes the same $(8, 4)$ binary linear block code;  note that the variable nodes of Figure 1(a) have been replaced by equality constraints, so that the internal variables are replicas of the external variables.  The code $\CC$ realized by the graph consists of all external variable assignments $\ab$ such that there exists an internal variable assignment $\sb$ such that all local constraints are satisfied.

\subsection{Normal realizations and normal factor graphs}

In this paper, we will mostly represent codes (and weight generating
functions of codes) by normal factor graphs and their partition functions.
A \emph{partition function} will be defined by
\begin{itemize}
\item a set $\Ab = \{A_k, k \in \I_\A\}$ of \emph{external variables} $A_k$
with alphabets $\A_k$;
\item a set $\Sb = \{S_j, j \in \I_\SSS\}$ of \emph{internal variables}
$S_j$ with alphabets $\SSS_j$;
\item a set $\fb = \{f_i, i \in \I_\FF\}$ of complex-valued \emph{factors}
$f_i$, each factor $f_i$ involving subsets $\Ab_i \subseteq \Ab$ and $\Sb_i
\subseteq \Sb$ of the sets of internal and external variables.
\end{itemize}
All sets are assumed to be finite, and all variable alphabets are assumed to
be finite-dimensional vector spaces over some finite field $\F$.  We call
the Cartesian-product alphabet $\A = \prod_k \A_k$ the \emph{external
variable configuration space}, and $\SSS = \prod_j \SSS_j$ the
\emph{internal variable configuration space}.

The partition function\footnote{In physics, a partition function is usually
defined as a sum over internal configurations, and there are no external
variables, so this usage of ``partition function" extends the usual
terminology of physics.}
 (or ``external function" \cite{AM10}) defined by these elements is the
function $Z: \A \to \C$ of the external variables that is given by the
following sum of products:
$$
Z(\ab) = \sum_{\sb \in \SSS} \prod_{i \in \I_\FF} f_i(\ab_i, \sb_i), \quad
\ab \in \A;
$$
\ie the sum over all  internal variable configurations of
the product of all  factors.

A partition function $Z(\ab)$ may in general be given by many
different sum-of-products forms, which we will call \emph{realizations}.  We will say that
two realizations of the same partition function are \emph{equivalent}.

A realization will be called \emph{normal} if each external variable is
involved in precisely one factor, and each internal variable is involved in
precisely two factors.
As noted in \cite{F01}, any realization may be converted to an equivalent normal realization by the following simple \textbf{normalization procedure}:
\begin{itemize}
\item For every external variable $A_i$, if $A_i$ is involved in $p$ factors, then define $p$ \emph{replica variables} $A_{i\ell}, 1 \le \ell \le p$, replace $A_i$ by $A_{i\ell}$ in the $\ell$th factor in which $A_i$ is involved, and introduce one new factor, namely an \emph{equality indicator function} $\Phi_{=}(a_i, \{a_{i\ell}, 1 \le \ell \le p\})$;  \ie the $(0,1)$-valued function that equals 1 when $a_i = a_{i1} = \cdots = a_{ip}$, and that equals 0 otherwise.  Thus each replica variable $A_{i\ell}$ becomes an internal variable that is involved in precisely two factors, while $A_i$ remains an external variable that is involved in only one factor, namely the equality indicator function.  (If $p= 1$, then this conversion need not be performed.)
\item For every internal variable $S_j$, if $S_j$ is involved in $q \ge 2$ factors, then define $q$ replica variables $S_{j\ell}, 1 \le \ell \le q$, replace $S_j$ by $S_{j\ell}$ in the $\ell$th factor in which $S_j$ is involved, and introduce one new factor, namely an equality indicator function $\Phi_{=}(\{s_{j\ell}, 1 \le \ell \le q\})$.  Thus each replica variable $S_{j\ell}$ becomes an internal variable that is involved in precisely two factors.  (If $q=2$, then this conversion need not be performed.  If $q=1$, then multiply the partition function by a dummy factor $1(s_k)$ which is equal to 1 regardless of the value of $S_k$.)
\end{itemize}
Evidently this normalization procedure does not change the partition function $Z(\ab)$.  Also, as can be seen from \cite{F01}, the normal factor graph that represents the normal realization is essentially unchanged from the bipartite factor graph that represents the original realization.

A normal realization is represented by a \emph{normal factor graph} (NFG).\footnote{Loeliger \etal \cite{L04, L07} define ``Forney-style factor graphs" just as we have defined normal factor graphs.  However, in the usual factor graph framework, such a graph represents simply the product of the factors, rather than a sum of products.  Loeliger \etal do also consider sums of products within ``boxes," namely graph fragments  enclosed by dashed lines.} As in a normal graph, ordinary edges represent internal variables and half-edges represent external variables, but now vertices represent factors rather than constraints.  The partition function of the NFG is the partition function $Z(\ab)$ of the associated realization.  

In this paper we will mostly adopt the convention of ignoring multiplicative scale factors $\alpha > 0$ in partition functions.  We will say that the partition function is equal to $Z(\ab)$, \emph{up to scale};  that is, the relative weights $Z(\ab)$ for the various configurations $\ab \in \A$ are correct, but the absolute value may not be.  In many applications the absolute scale factor is not important; but if it is, then it can be reconstructed from the constituent factors $f_i$.

A normal graph representing a code $\CC$ may be converted to a normal factor graph representing the \emph{indicator function} $\Phi_\CC$ of $\CC$ as follows.  Let each local constraint code $\CC_i$ be replaced by the $(0,1)$-valued indicator function $\Phi_{\CC_i}$ of $\CC_i$.  Then, for a given external variable assignment $\ab \in \A$, the partition function of the graph--- \ie the sum over all $\sb \in \SSS$ of the product of all local constraint code indicator functions--- is the number of internal variable assignments $\sb \in \SSS$ such that $(\ab, \sb)$ satisfies all local constraints.  For a linear code $\CC$, it is often true that a unique $\sb \in \SSS$ is determined by each $\ab \in \CC$; \eg when the graph is cycle-free and the realization is minimal.  But in any case, by linearity, the same number of state sequences correspond to every $\ab \in \CC$, namely $|\{\sb \in \SSS: (\zerob, \sb) \in \Bf\}|$.  
Therefore, up to scale, the partition function of the NFG is equal to the $(0,1)$-valued indicator function $\Phi_\CC$;  \ie
$$
\Phi_\CC(\ab) \propto \sum_{\sb \in \SSS} \prod_{i \in \I_\CC} \Phi_{\CC_i}(\ab_i, \sb_i), \quad \ab \in \A.
$$

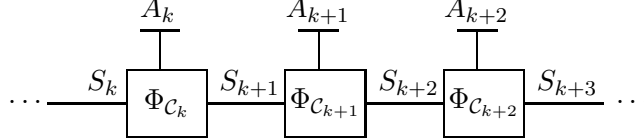
\begin{figure}[h]
\setlength{\unitlength}{5pt}
\centering
\begin{picture}(50,7)(-2, 4)
\multiput(1,5)(12,0){4}{\line(1,0){6}}
\multiput(9.5,7.5)(12,0){3}{\line(0,1){3}}
\multiput(8,10.5)(12,0){3}{\line(1,0){3}}
\put(-2,5){\ldots}
\put(4,6){$S_k$}
\put(14,6){$S_{k+1}$}
\put(26,6){$S_{k+2}$}
\put(38,6){$S_{k+3}$}
\put(44,5){\ldots}
\put(8,11.5){$A_k$}
\put(19,11.5){$A_{k+1}$}
\put(31,11.5){$A_{k+2}$}
\put(7,2.5){\framebox(6,5){$\Phi_{\CC_k}$}}
\put(19,2.5){\framebox(6,5){$\Phi_{\CC_{k+1}}$}}
\put(31,2.5){\framebox(6,5){$\Phi_{\CC_{k+2}}$}}
\end{picture}
\caption{Normal factor graph of a conventional state realization.}
\label{Fig2}
\end{figure}

For example, Figure \ref{Fig2} shows the normal factor graph corresponding to the normal graph of Figure \ref{Fig1}.

%

%\pagebreak
\subsection{Weight generating functions}

Weight generating functions are standard tools of combinatorics.  We will later consider various kinds of weight generating functions, but for linear codes over finite fields, the simplest and commonest kind of weight generating functions are Hamming weight generating functions.  

The \emph{Hamming weight generating function} (HWGF) of a linear code $\CC$ defined over a finite index set $\I_\A$ with symbol sequence space $\A = \prod_{k \in \I_\A} A_k$ is the polynomial
$$
g^H_\CC(x) = \sum_{\ab \in \CC} \prod_{k \in \I_\A} x^{w_H(a_k)},
$$
where $w_H(a_k)$ is the Hamming weight of the symbol $a_k \in A_k$.  Thus the coefficient of $x^w$ in $g^H_\CC(x)$ is the number of codewords $\ab \in \CC$ that have Hamming weight $w$.

\begin{figure}[h]
\setlength{\unitlength}{5pt}
\centering
\begin{picture}(70,9)(-2, 2)
\put(13,5){\line(1,0){8}}
\put(16,6){$\A$}
\put(7,2.5){\framebox(6,5){$\Phi_{\CC}(\ab)$}}
\put(21,2.5){\framebox(8,5){$x^{w_H(\ab)}$}}
\put(16,1){(a)}
\put(37,2.5){\framebox(6,7){$\Phi_{\CC}(\ab)$}}
\put(43,9){\line(1,0){14}}
\put(46,10){$A_1$}
\put(57,7.5){\framebox(10,3){$x^{w_H(a_1)}$}}
\put(43,3){\line(1,0){14}}
\put(46,4){$A_{|\I_\A|}$}
\put(46,5.5){$\cdot$}
\put(46,6.5){$\cdot$}
\put(46,7.5){$\cdot$}
\put(57,1.5){\framebox(10,3){$x^{w_H(a_{|\I_\A|})}$}}
\put(48,1){(b)}
\end{picture}
\caption{Normal factor graphs of the HWGF $g^H_\CC(x)$ of a code $\CC \subseteq \A$.}
\label{Fig3}
\end{figure}
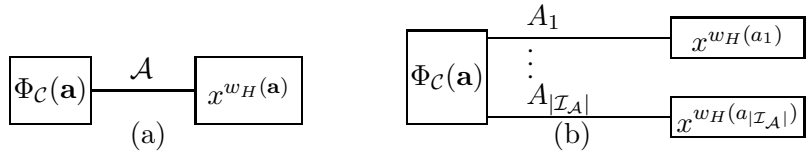

This sum-of-products expression for $g^H_\CC(x)$ suggests that a HWGF might be represented as the partition function of a normal factor graph.    (This idea was briefly mentioned in \cite[Example 13]{KFL01}.)  Indeed, since
$$
g^H_\CC(x) = \sum_{\ab \in \A} \Phi_{\CC}(\ab) x^{w_H(\ab)},
$$
it follows that $g^H_\CC(x)$ is the partition function of the simple normal factor graph of Figure \ref{Fig3}(a), in which the two functions $\Phi_\CC(\ab)$ and $x^{w_H(\ab)}$ are connected by the internal variable $\A$.

  Or, since $w_H(\ab) = \sum_{k \in \I_\A} w_H(a_k)$ is an additive function, $g^H_\CC(x)$ may alternatively be represented as the partition function of the normal factor graph of Figure \ref{Fig3}(b), in which the function $\Phi_\CC(\ab)$  is connected by an internal variable $A_k$ to the function $x^{w_H(a_k)}$ for each $k \in \I_\A$.
  
Now, more generally, if the indicator function $\Phi_\CC$ of a linear code $\CC$ is the partition function of any normal factor graph, then a normal factor graph for $g^H_\CC(x)$ may be obtained by replacing the symbol half-edge associated with $A_k$ in that graph by an ordinary edge connected to the function $x^{w_H(a_k)}$ of $A_k$ for each $k \in \I_\A$.  (Or, if $w_H(a_k) = \sum_{\ell} w_H(a_{k\ell})$ for some components $a_{k\ell}$ of $a_k$, then $A_k$ and $x^{w_H(a_k)}$ may be broken down into their components $A_{k\ell}$ and $x^{w_H(a_{k\ell})}$ as in Figure \ref{Fig3}(b).)  

\begin{figure}[h]
\setlength{\unitlength}{5pt}
\centering
\begin{picture}(50,12)(-2, 3)
\multiput(1,5)(12,0){4}{\line(1,0){6}}
\multiput(10,7.5)(12,0){3}{\line(0,1){4}}
\put(-2,5){\ldots}
\put(4,6){$S_k$}
\put(14,6){$S_{k+1}$}
\put(26,6){$S_{k+2}$}
\put(38,6){$S_{k+3}$}
\put(44,5){\ldots}
\put(5,11.5){\framebox(10,3){$x^{w_H(a_k)}$}}
\put(17,11.5){\framebox(10,3){$x^{w_H(a_{k+1})}$}}
\put(29,11.5){\framebox(10,3){$x^{w_H(a_{k+2})}$}}
\put(10.3,9){$A_{k}$}
\put(22.3,9){$A_{k+1}$}
\put(34.3,9){$A_{k+2}$}
\put(7,2.5){\framebox(6,5){$\Phi_{\CC_k}$}}
\put(19,2.5){\framebox(6,5){$\Phi_{\CC_{k+1}}$}}
\put(31,2.5){\framebox(6,5){$\Phi_{\CC_{k+2}}$}}
\end{picture}
\caption{Normal factor graph of the HWGF $g^H_\CC(x)$ of the code $\CC$ of Figure 2.}
\label{Fig4}
\end{figure}
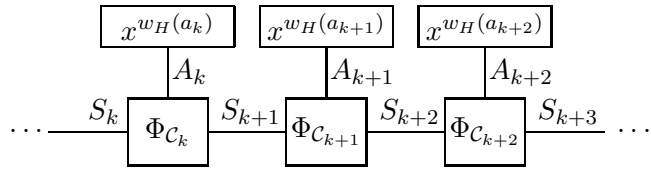

%\pagebreak
For example, for a normal factor graph of a conventional state realization as in Figure \ref{Fig2}, the HWGF $g^H_\CC(x)$ is the partition function of the normal factor graph shown in Figure \ref{Fig4}.

Such a representation  of $g^H_\CC(x)$ may be further simplified by summing over the variables $A_k$ for each $k \in \I_\A$, which have now become internal variables.  For a code represented by a conventional state realization as in Figure \ref{Fig4}, this results in a normal factor graph as in Figure \ref{Fig5}.  Here $\Lambda_{\CC_k}^H(s_k, s_{k+1})$ is what is called \cite{GLS08, GLS09} the \emph{Hamming weight adjacency matrix}\footnote{Perhaps this should have been called a ``Hamming weight generating function adjacency matrix," but we prefer to follow the established terminology.} (HWAM) of $\CC_k$:
$$
\Lambda_{\CC_k}^H(s_k, s_{k+1})(x) = \sum_{a_k \in A_k} \Phi_{\CC_k}(s_k, a_k, s_{k+1}) x^{w_H(a_k)}
$$

\begin{figure}[h]
\setlength{\unitlength}{5pt}
\centering
\begin{picture}(60,3)(-2, 4)
\multiput(1,5)(15,0){4}{\line(1,0){6}}
\put(-2,5){\ldots}
\put(4,6){$S_k$}
\put(17,6){$S_{k+1}$}
\put(32,6){$S_{k+2}$}
\put(47,6){$S_{k+3}$}
\put(53,5){\ldots}
\put(7,2.5){\framebox(9,5){$\Lambda^H_{\CC_k}(x)$}}
\put(22,2.5){\framebox(9,5){$\Lambda^H_{\CC_{k+1}}(x)$}}
\put(37,2.5){\framebox(9,5){$\Lambda^H_{\CC_{k+2}}(x)$}}
\end{picture}
\caption{Normal factor graph resulting from summing over each $A_k$ in Figure 4.}
\label{Fig5}
\end{figure}

%\pagebreak

In equations, the equivalence of the partition functions of Figures \ref{Fig4} and \ref{Fig5} follows from
\begin{eqnarray*}
g^H_\CC(x) & \propto & \sum_{\sb \in \SSS} \sum_{\ab \in \A} \prod_{k \in \I_\A} \Phi_{\CC_k}(s_k, a_k, s_{k+1}) x^{w_H(a_k)} \\
& = & \sum_{\sb \in \SSS} \prod_{k \in \I_\A} \sum_{a_k \in A_k}  \Phi_{\CC_k}(s_k, a_k, s_{k+1}) x^{w_H(a_k)} \\
& = & \sum_{\sb \in \SSS} \prod_{k \in \I_\A} \Lambda_{\CC_k}^H(s_k, s_{k+1})(x).
\end{eqnarray*}
The last expression may be recognized as simply the product $\cdots  \Lambda_{\CC_k}(x)  \Lambda_{\CC_{k+1}}(x)  \Lambda_{\CC_{k+2}}(x) \cdots$ of the HWAMs, using the usual rules of matrix arithmetic.

\begin{figure}[h]
\setlength{\unitlength}{5pt}
\centering
\begin{picture}(42,15)(0, 0)
\put(0,6){\framebox(2,2){$0$}}
\put(2,7){\line(4,3){8}}
\put(2,7){\line(4,1){8}}
\put(2,7){\line(4,-1){8}}
\put(2,7){\line(4,-3){8}}
\put(5,11){$00$}
\put(5,8){$11$}
\put(5,5){$01$}
\put(5,2){$10$}
\put(10,0){\framebox(2,2){$11$}}
\put(10,4){\framebox(2,2){$01$}}
\put(10,8){\framebox(2,2){$10$}}
\put(10,12){\framebox(2,2){$00$}}
\put(20,0){\framebox(2,2){$11$}}
\put(20,4){\framebox(2,2){$01$}}
\put(20,8){\framebox(2,2){$10$}}
\put(20,12){\framebox(2,2){$00$}}
\put(12,1){\line(1,0){8}}
\put(12,1){\line(2,1){8}}
\put(12,5){\line(2,-1){8}}
\put(12,13){\line(1,0){8}}
\put(12,9){\line(2,1){8}}
\put(12,13){\line(2,-1){8}}
\put(12,9){\line(1,0){8}}
\put(12,5){\line(1,0){8}}
\put(15,9){$00$}
\put(13,11){$11$}
\put(15,13){$00$}
\put(17,11){$11$}
\put(15,1){$01$}
\put(13,3){$10$}
\put(15,5){$01$}
\put(17,3){$10$}
\put(30,0){\framebox(2,2){$11$}}
\put(30,4){\framebox(2,2){$01$}}
\put(30,8){\framebox(2,2){$10$}}
\put(30,12){\framebox(2,2){$00$}}
\put(22,1){\line(1,0){8}}
\put(22,1){\line(2,1){8}}
\put(22,5){\line(2,-1){8}}
\put(22,13){\line(1,0){8}}
\put(22,9){\line(2,1){8}}
\put(22,13){\line(2,-1){8}}
\put(22,9){\line(1,0){8}}
\put(22,5){\line(1,0){8}}
\put(25,9){$00$}
\put(23,11){$11$}
\put(25,13){$00$}
\put(27,11){$11$}
\put(25,1){$01$}
\put(23,3){$10$}
\put(25,5){$01$}
\put(27,3){$10$}
\put(40,6){\framebox(2,2){$0$}}
\put(32,13){\line(4,-3){8}}
\put(32,9){\line(4,-1){8}}
\put(32,5){\line(4,1){8}}
\put(32,1){\line(4,3){8}}
\put(35,11){$00$}
\put(35,8){$11$}
\put(35,5){$01$}
\put(35,2){$10$}

\end{picture}

\caption{Four-section state realization of $(8, 4)$ binary first-order Reed-Muller code.}
\label{Fig6}
\end{figure}
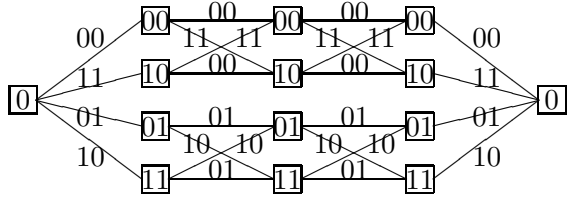

\vspace{1ex}
%\pagebreak
\noindent
\textbf{Example 1} (binary linear block code).  Consider the $(8, 4)$ binary linear first-order Reed-Muller code $\CC$, which has the conventional four-section state realization (trellis) shown in Figure \ref{Fig6} \cite{F01}.

\pagebreak
\noindent
The Hamming weight generating function $g^H_\CC(x)$ is given by the product of the HWAMs of each of the four sections, as follows:
$$
\left[
\begin{array}{cccc}
1 & x^2 & x & x
\end{array}
\right]
\left[
\begin{array}{cccc}
1 & x^2 & 0 & 0 \\
x^2 & 1 & 0 & 0 \\
0 & 0 & x & x \\
0 & 0 & x & x
\end{array}
\right]
\left[
\begin{array}{cccc}
1 & x^2 & 0 & 0 \\
x^2 & 1 & 0 & 0 \\
0 & 0 & x & x \\
0 & 0 & x & x
\end{array}
\right]
\left[
\begin{array}{c}
1 \\ x^2 \\ x \\ x
\end{array}
\right]
 = 1 + 14x^2 + x^8.  \qed
$$
%\vspace{1ex}

Let us now consider the Hamming weight generating function of a fragment of a normal factor graph realizing a code $\CC$, consisting of a subset of the constraint codes $\CC_i$.  An external (symbol) variable $A_k$ may or may not be involved in one of these constraint codes;  if it is, we attach an enumerator function $x^{w_H(a_k)}$ as above.   An internal (state) variable $S_j$ may appear in two, one, or none of these constraint codes:  if it appears twice, then we consider it to be an internal variable of the fragment and sum over it;\footnote{Sometimes physicists use the Einstein summation convention: in a tensor product, variables that occur twice are implicitly to be summed over.  This convention would evidently be useful for partition functions of normal factor graphs.}  if it appears once, then we consider it to be an external variable of the fragment.  In other words,  the HWGF of the fragment is the partition function
$$
g^H_f(x)(\sb_{\mathrm{ext}}) \propto \sum_{\sb_{\mathrm{int}}} \prod_{i \in \I_{\A,f}} \sum_{\ab_i} \Phi_{\CC_i}(\ab_i, \sb_i) x^{w_H(\ab_i)},
$$
where $\sb_{\mathrm{ext}}$ and $\sb_{\mathrm{int}}$ are the external and internal state variables with respect to the fragment, and $\I_{\A,f} \subseteq \I_\A$ is the subset of indices of constraint codes $\CC_i$ that appear in the fragment.

\vspace{1ex}
\noindent
\textbf{Example 2} ($N$ sections of a linear convolutional code).  Consider a linear time-invariant convolutional code $\CC$, in which all symbol alphabets $A_k$, state spaces $S_k$, and constraint codes $\CC_k$ do not actually depend on the time index $k \in \Z$.  Then the HWAM $\Lambda^H_{\CC_k}(x)$ does not depend upon $k$, so we write it simply as $\Lambda(x)$.  Consider a fragment of a graph for $\CC$ consisting of $N$ consecutive trellis sections, over a time interval $[k, k+N)$.  Then, relative to this fragment, the external variables are $S_k$ and $S_{k+N}$, and the HWAM of the fragment is simply the $N$th power $\Lambda^N(x)$ of the HWAM $\Lambda(x)$ of a single section.  \qed \vspace{1ex}

\section{Duality}

The central duality result of \cite{F01} is the \emph{normal graph duality theorem}:  given a normal realization of a code $\CC$, the dual normal realization generates the dual (orthogonal) code $\CC^\perp$.  We will now give a concise proof of this result, as a corollary of a general duality theorem for normal factor graphs.  

\subsection{Fourier transforms}

We first recall the theory of Fourier transforms over finite fields, generally following \cite{F98}.

Let the prime $p$ be the characteristic of the finite field $\F$;  \ie $p$ is the least positive integer such that $p\alpha = 0$ for all $\alpha \in \F$.

If $A$ is a vector space over $\F$, then its dual space $\hat{A}$ may be defined as the set of all homomorphisms $\hat{a}:  A \to \Z_p$.  It follows from Pontryagin duality theory \cite{F98} that $\hat{A}$ is a vector space over $\F$ of the same dimension as $A$, and that the dual space to $\hat{A}$ is $A$, with $a(\hat{a})$ defined as $\hat{a}(a)$.  We may then define the \emph{inner product} $\inner{\hat{a}}{a}$ as $a(\hat{a}) = \hat{a}(a)$ for all $a \in A, \hat{a} \in \hat{A}$.  The inner product so defined has the usual properties;  \eg $\inner{\hat{a}}{0} = \inner{0}{a} = 0$, $\inner{\hat{a}}{a + a'} = \inner{\hat{a}}{a} + \inner{\hat{a}}{a'}$, and so forth.  

For example, if $A$ is the field $\F$, then the additive group of $\F$ is isomorphic to $(\Z_p)^m$ for some integer $m$.  The dual space $\hat{A}$ may also be taken as $\F \simeq (\Z_p)^m$, with the inner product defined componentwise as the dot product $\hat{a} \cdot {a} = \sum_{i = 1}^m \hat{a}_i a_i,$ with all operations in the prime field $\Z_p$.

Similarly, if $A$ is the set $\F^n$ of all $n$-tuples over $\F$, then its dual space $\hat{A}$ may also be taken as $\F^n$, and the inner product may again be defined componentwise as 
$$\inner{\hat{a}}{a} = \sum_{j = 1}^n \hat{a}_j \cdot a_j= \sum_{j = 1}^n \sum_{i = 1}^m \hat{a}_{ji} a_{ji} $$

Given a complex-valued function $f: A \to \C$ defined on $A$, its \emph{Fourier transform} is defined as the complex-valued function $F: \hat{A} \to \C$ that maps $\hat{a} \in \hat{A}$ to
$$
F(\hat{a}) = \sum_{a \in A} f(a)  \omega^{\inner{\hat{a}}{a}}, \quad \hat{a} \in \hat{A},
$$
where $\omega = e^{2\pi i/p}$ is a primitive complex $p$th root of unity. 

 If we view $\fb = \{f(a) : a \in A\}$ as a column vector indexed by $A$, and similarly $\Fb = \{F(\hat{a}) : \hat{a} \in \hat{A}\}$ as a column vector indexed by $\hat{A}$, then the transform can be expressed in matrix form as
$$
\Fb = \FF_A \fb,
$$
where the \emph{Fourier transform matrix} $\FF_A$ is defined as $\{\omega^{\inner{\hat{a}}{a}}: \hat{a} \in \hat{A}, a \in A\}$.  Note that $\FF_A$ is symmetric;  \ie $\FF_A^T = \FF_A$, where $\FF_A^T$ denotes the transpose of $\FF_A$.  %Furthermore, $\FF_A^T = \{\omega^{\inner{\hat{a}}{a}}: a \in A, \hat{a} \in \hat{A}\}$ is the transform matrix $\FF_{\hat{A}}$.

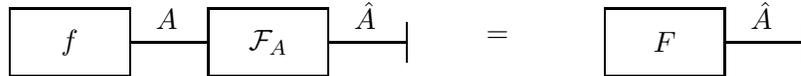
\begin{figure}[h]
\setlength{\unitlength}{5pt}
\centering
\begin{picture}(70,5)(0, 3)
\multiput(16,5)(15,0){2}{\line(1,0){6}}
\put(37,3.5){\line(0,1){3}}
\put(18,6){$A$}
\put(33,6){$\hat{A}$}
\put(7,2.5){\framebox(9,5){$f$}}
\put(22,2.5){\framebox(9,5){$\FF_A$}}
\put(43,5){$=$}
\put(52,2.5){\framebox(9,5){$F$}}
\put(61,5){\line(1,0){6}}
\put(67,3.5){\line(0,1){3}}
\put(63,6){$\hat{A}$}
\end{picture}
\caption{Normal factor graph of a Fourier transform.}
\label{Fig7}
\end{figure}

In a normal factor graph, a Fourier transform may be simply represented as in Figure \ref{Fig7}.  The transform $F(\hat{a})$ is obtained by summing over $A$, which in this case amounts to a matrix multiplication.  Note that as a factor in a factor graph, we do not have to distinguish between $\FF_A$ and its transpose;  $\FF_A$ is simply a function of the two variables corresponding to the two incident edges, and as a matrix can act on either variable.

From the \emph{orthogonality relation} 
$$
\sum_{\hat{a} \in \hat{A}}  \omega^{\inner{\hat{a}}{a}} = \left\{
\begin{array}{cc}
 |A|,  & a = 0; \\ 
 0, & a \neq 0, 
 \end{array} \right.
$$
we obtain the matrix equation 
$$\FF_A^* \FF_A = |A| I_A,$$
 where $\FF_A^* = \{\omega^{-\inner{\hat{a}}{a}} : a \in A, \hat{a} \in \hat{A}\}$ is the conjugate transpose of $\FF_A$, and $I_A$ is the identity matrix over $A$.  In other words, the inverse of $\FF_A$ is $\FF_A^{-1} = |A|^{-1} \FF_A^*$.
Thus we obtain the \emph{inverse Fourier transform}
$$
\fb = \FF_A^{-1} \Fb = \frac{\FF_A^*\Fb}{|A|}.
$$
We say that $\fb$ and $\Fb$ are a \emph{Fourier transform pair}, written $\fb \lra \Fb$. %\pagebreak

More generally, we have the following orthogonality relation (which even more generally applies to orthogonal subgroups of finite abelian groups):

\vspace{1ex}
\noindent
\textbf{Orthogonality relation for subspaces}.  Let $B$ be any subspace of a finite subspace $A$, and let $B^\perp$ be the orthogonal subspace of its dual space $\hat{A}$;  \ie $B^\perp$ is the subset of elements $\hat{a} \in \hat{A}$ such that $\inner{\hat{a}}{a} = 0$ for all $a \in B$.  Then
$$
\sum_{a \in B}  \omega^{\inner{\hat{a}}{a}} = \left\{
\begin{array}{cc}
 |B|,  & \hat{a} \in B^\perp; \\ 
 0, & \hat{a} \notin B^\perp.
 \end{array} \right.
$$
\textit{Proof}:  Obvious for $\hat{a} \in B^\perp$.  For $\hat{a} \notin B^\perp$, let $z(\hat{a}) = \sum_{a \in B}  \omega^{\inner{\hat{a}}{a}}$, and let $a'$ be some element of $B$ such that $\inner{\hat{a}}{a'} \neq 0$.  Since $a' + B = B$, we have $z(\hat{a}) = \sum_{a \in B}  \omega^{\inner{\hat{a}}{a' + a}} = \omega^{\inner{\hat{a}}{a'}} z(\hat{a})$;  but $\inner{\hat{a}}{a'} \neq 0$ implies $\omega^{\inner{\hat{a}}{a'}} \neq 1$, so $z(\hat{a}) =   \omega^{\inner{\hat{a}}{a'}} z(\hat{a})$ implies $z(\hat{a}) = 0$. \qed \vspace{1ex}

Let $\Phi_B:  A \to \{0,1\}$ be the indicator function of the subspace $B$;  \ie $\Phi_B(a) = 1$ if $a \in B$, else $\Phi_B(a) = 0$.  From this orthogonality relation, it follows that the Fourier transform of $\Phi_B$ is $|B| \Phi_{B^\perp}$.  Conversely, by a similar argument, the inverse Fourier transform of $\Phi_{B^\perp}$ is $|B^\perp | \Phi_B/ |A|$.  Thus if $B$ and $B^\perp$ are orthogonal subspaces of a finite vector space $A$ and its dual space $\hat{A}$, respectively, then
\begin{itemize}
\item the indicator functions $\Phi_B$ and $\Phi_{B^\perp}$ are a Fourier transform pair, up to scale; 
\item $|B||B^\perp| = |A| = |\hat{A}|$.
\end{itemize}
As  \cite{MK05} notes, the top result is a version of the ``picket-fence miracle" \cite{F98} (see also \cite[Theorem 9]{L07}).

We observe that the Poisson summation formula, namely
$$
\sum_{\ab \in B} f(\ab) = |B^\perp| \sum_{\hat{\ab} \in B^\perp} F(\hat{\ab}),
$$
follows directly from the Fourier transform relationship for orthogonal code indicator functions, where the functions $f \lra F$ are any Fourier transform pair;  see Figure \ref{Fig10}.  (Note that since both $\Phi_B$ and $\Phi_{B^\perp}$ are real-valued, the Fourier transform relation holds with the kernel $\FF_A^*$ as well as with $\FF_A$.) We conclude that the Fourier transform pair relationship $\Phi_B \lra \Phi_{B^\perp}$ is more fundamental than the Poisson summation formula, which is the usual starting point in the development of MacWilliams identities.

\begin{figure}[h]
\setlength{\unitlength}{5pt}
\centering
\begin{picture}(70,3)(8, 7)
\put(10,6){\framebox(5,3){$\Phi_B$}}
\put(15,7.5){\line(1,0){5}}
\put(17,8){$A$}
\put(20,6){\framebox(5,3){$f$}}
\put(27,7){$=$}
\put(30,6){\framebox(5,3){$\Phi_B$}}
\put(35,7.5){\line(1,0){5}}
\put(37,8){$A$}
\put(40,6){\framebox(5,3){$\FF_A^*$}}
\put(45,7.5){\line(1,0){5}}
\put(47,8){$\hat{A}$}
\put(50,6){\framebox(5,3){$F$}}
\put(57,7){$=$}
\put(60,6){\framebox(5,3){$\Phi_{B^\perp}$}}
\put(65,7.5){\line(1,0){5}}
\put(67,8){$\hat{A}$}
\put(70,6){\framebox(5,3){$F$}}
\end{picture}
\caption{Proof of the Poisson summation formula.}
\label{Fig10}
\end{figure}

\begin{figure}[h]
\setlength{\unitlength}{5pt}
\centering
\begin{picture}(70,4)(0, 3)
\multiput(16,5)(15,0){2}{\line(1,0){6}}
\put(37,3.5){\line(0,1){3}}
\put(18,6){$A$}
\put(33,6){$\hat{A}$}
\put(7,2.5){\framebox(9,5){$x$}}
\put(22,2.5){\framebox(9,5){$\FF_A$}}
\put(43,5){$=$}
\put(52,2.5){\framebox(9,5){$X$}}
\put(61,5){\line(1,0){6}}
\put(67,3.5){\line(0,1){3}}
\put(63,6){$\hat{A}$}
\end{picture}
\caption{Normal factor graph of a set $\xb$ of indeterminates indexed by $A$, and the dual set $\Xb$.}
\label{Fig8}
\end{figure}
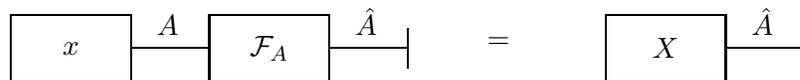

Finally, we extend these definitions to a set of indeterminates $\xb = \{x(a) : a \in A\}$ indexed by $A$, rather than a complex-valued function.  The transform of this set is then a dual set of indeterminates $\Xb = \{X(\hat{a}) : \hat{a} \in \hat{A}\}$ indexed by $\hat{A}$, defined by
$$
\Xb = \FF_A \xb.
$$
Again, we have the inverse transform relationship
$$
\xb = \FF_A^{-1} \Xb = \frac{\FF_A^*\Xb}{|A|},
$$
and we say that $\xb$ and $\Xb$ are a transform pair, written $\xb \lra \Xb$.   Figure \ref{Fig8} shows the corresponding normal factor graph, similar to Figure \ref{Fig7}.  %\pagebreak

For example, if $A = \Z_2$, then $X(0) = x(0) + x(1)$ and $X(1) = x(0) - x(1)$;  similarly, $x(0) = \half(X(0) + X(1))$, and $x(1) = \half(X(0) -X(1))$.  For another example, if $A = (\Z_2)^2$, then 
\begin{eqnarray*}
X(00) &=& x(00) + x(10) + x(01) + x(11) \\ X(10) &=& x(00) - x(10) + x(01) - x(11) \\ X(01) &=& x(00) + x(10) - x(01) - x(11) \\ X(11) &=& x(00) - x(10) - x(01) + x(11)
\end{eqnarray*}
and \emph{vice versa} (up to a scale factor of $\frac{1}{4}$).

\subsection{Dual linear codes}

Again, a linear code $\CC$ over a finite field $\F$ is a subspace $\CC \subseteq \A$ of a symbol sequence space $\A = \prod_{k \in \I_\A} A_k$, where each symbol alphabet $A_k$ is a finite-dimensional vector space over $\F$.  

As we have seen, each vector space $A_k$ has a dual space $\hat{A}_k$ of the same dimension such that for all $a_k \in A_k, \hat{a}_k \in \hat{A}_k$ there is a well-defined inner product $\inner{\hat{a}_k}{a_k}$.  The dual space to the symbol sequence space $\A = \prod_{k \in \I_\A} A_k$ is then the dual symbol sequence space $\hat{\A} = \prod_{k \in \I_\A} \hat{A}_k$, with the inner product defined componentwise by $\inner{\hat{\ab}}{\ab} = \sum_{k \in \I_\A} \inner{\hat{a}_k}{a_k}$ (where we continue to assume that $\I_\A$ is finite). 
The dual (orthogonal) code $\CC^\perp \subseteq \hat{\A}$ to $\CC \subseteq \A$ is then the set of all dual symbol sequences $\hat{\ab} \in \hat{\A}$ such that $\inner{\hat{\ab}}{\ab} = 0$ for all $\ab \in \CC$. 

From our general orthogonality theorem, we have that the indicator functions $\Phi_\CC$ and $\Phi_{\CC^\perp}$ are a Fourier transform pair, up to scale, and that $|\CC||\CC^\perp| = |\A|$.

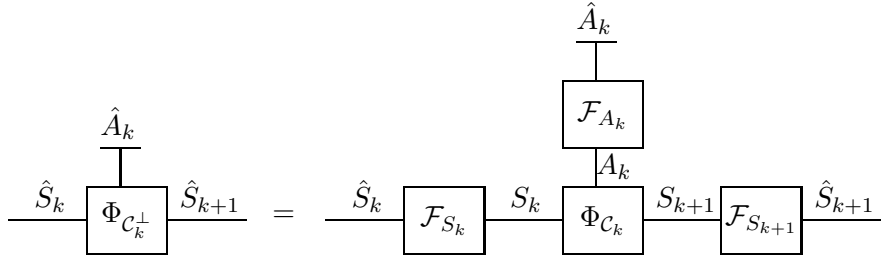
\begin{figure}[h]
\setlength{\unitlength}{5pt}
\centering
\begin{picture}(70,13)(-4, 3)
\multiput(1,5)(12,0){2}{\line(1,0){6}}
\multiput(9.5,7.5)(12,0){1}{\line(0,1){3}}
\multiput(8,10.5)(12,0){1}{\line(1,0){3}}
\put(3,6){$\hat{S}_k$}
\put(14,6){$\hat{S}_{k+1}$}
\put(8,11.5){$\hat{A}_k$}
\put(7,2.5){\framebox(6,5){$\Phi_{\CC_k^\perp}$}}
\put(21,5){$=$}
\multiput(25,5)(12,0){4}{\line(1,0){6}}
\multiput(45.5,7.5)(0,8){2}{\line(0,1){3}}
\multiput(44,18.5)(12,0){1}{\line(1,0){3}}
\put(27,6){$\hat{S}_k$}
\put(39,6){$S_k$}
\put(50,6){$S_{k+1}$}
\put(62,6){$\hat{S}_{k+1}$}
\put(44,19.5){$\hat{A}_k$}
\put(45.5,8.5){$A_k$}
\put(31,2.5){\framebox(6,5){$\FF_{S_k}$}}
\put(43,2.5){\framebox(6,5){$\Phi_{\CC_k}$}}
\put(55,2.5){\framebox(6,5){$\FF_{S_{k+1}}$}}
\put(43,10.5){\framebox(6,5){$\FF_{A_k}$}}
\end{picture}
\caption{Transform of constraint code indicator function $\Phi_{\CC_k}$ in a conventional state realization.}
\label{Fig9}
\end{figure}

Since the inner product is defined componentwise, \ie $\inner{\hat{\ab}}{\ab} = \sum_{k \in \I_\A} \inner{\hat{a}_k}{a_k}$, it follows that the Fourier transform may be taken separately with respect to each variable $A_k$ involved in $\CC$.
For example, consider the indicator function $\Phi_{\CC_k}$ of a constraint code $\CC_k$ in a conventional state realization, as in Figure \ref{Fig2}.  The indicator function for the orthogonal code $\CC_k^\perp \subseteq \hat{S}_k \times \hat{A}_k \times \hat{S}_{k+1}$ may be obtained (up to scale) by transforming with respect to each of the three incident variables separately, as shown in the normal factor graph of Figure \ref{Fig9}.

\pagebreak
\subsection{Fourier transform identities}

In this subsection we develop a few quick and easy facts about the algebra of Fourier transforms.

We have already noted that $\FF_A \FF_A^* = I_A$, up to a scale factor of $|A|$.  We will write such a relationship as $\FF_A \FF_A^* \propto I_A$.  (Alternatively, we could normalize each Fourier transform matrix by an appropriate scale factor, namely $1/\sqrt{|A|}$ for $\FF_A$.)

We next observe that 
$$
\sum_{\hat{a} \in \hat{A}} \omega^{\inner{\hat{a}}{a}}\omega^{\inner{\hat{a}}{a'}} = \left\{
\begin{array}{cc}
 |A|,  & a + a' = 0; \\ 
 0, & a + a' \neq 0, 
 \end{array} \right.
$$
by the basic orthogonality relation.  In other words, $\FF_A^T \FF_A = |A|\Phi_{\sim A}$, where $\Phi_{\sim A}$ is the \emph{sign inverter indicator function} on $A$;  \ie the indicator function of the sign inversion relation, $a' = -a$.  

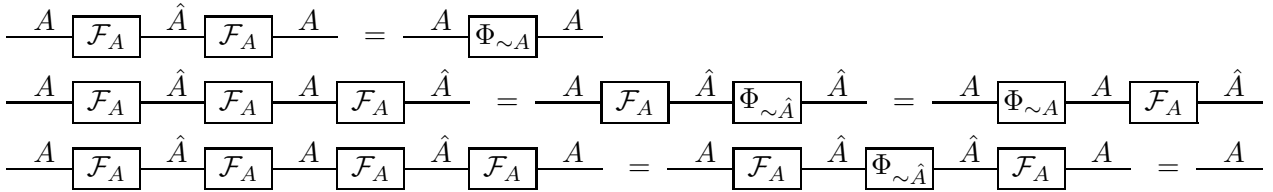
\begin{figure}[h]
\setlength{\unitlength}{5pt}
\centering
\begin{picture}(90,12)(8, 2)
\put(5,12.5){\line(1,0){5}}
\put(7,13){$A$}
\put(10,11){\framebox(5,3){$\FF_A$}}
\put(15,12.5){\line(1,0){5}}
\put(17,13){$\hat{A}$}
\put(20,11){\framebox(5,3){$\FF_A$}}
\put(25,12.5){\line(1,0){5}}
\put(27,13){$A$}
\put(32,12){$=$}
\put(35,12.5){\line(1,0){5}}
\put(37,13){$A$}
\put(40,11){\framebox(5,3){$\Phi_{\sim A}$}}
\put(45,12.5){\line(1,0){5}}
\put(47,13){$A$}
\put(5,7.5){\line(1,0){5}}
\put(7,8){$A$}
\put(10,6){\framebox(5,3){$\FF_A$}}
\put(15,7.5){\line(1,0){5}}
\put(17,8){$\hat{A}$}
\put(20,6){\framebox(5,3){$\FF_A$}}
\put(25,7.5){\line(1,0){5}}
\put(27,8){$A$}
\put(30,6){\framebox(5,3){$\FF_A$}}
\put(35,7.5){\line(1,0){5}}
\put(37,8){$\hat{A}$}
\put(42,7){$=$}
\put(45,7.5){\line(1,0){5}}
\put(47,8){$A$}
\put(50,6){\framebox(5,3){$\FF_A$}}
\put(55,7.5){\line(1,0){5}}
\put(57,8){$\hat{A}$}
\put(60,6){\framebox(5,3){$\Phi_{\sim \hat{A}}$}}
\put(65,7.5){\line(1,0){5}}
\put(67,8){$\hat{A}$}
\put(72,7){$=$}
\put(75,7.5){\line(1,0){5}}
\put(77,8){$A$}
\put(80,6){\framebox(5,3){$\Phi_{\sim A}$}}
\put(85,7.5){\line(1,0){5}}
\put(87,8){$A$}
\put(90,6){\framebox(5,3){$\FF_A$}}
\put(95,7.5){\line(1,0){5}}
\put(97,8){$\hat{A}$}
\put(5,2.5){\line(1,0){5}}
\put(7,3){$A$}
\put(10,1){\framebox(5,3){$\FF_A$}}
\put(15,2.5){\line(1,0){5}}
\put(17,3){$\hat{A}$}
\put(20,1){\framebox(5,3){$\FF_A$}}
\put(25,2.5){\line(1,0){5}}
\put(27,3){$A$}
\put(30,1){\framebox(5,3){$\FF_A$}}
\put(35,2.5){\line(1,0){5}}
\put(37,3){$\hat{A}$}
\put(40,1){\framebox(5,3){$\FF_A$}}
\put(45,2.5){\line(1,0){5}}
\put(47,3){$A$}
\put(52,2){$=$}
\put(55,2.5){\line(1,0){5}}
\put(57,3){$A$}
\put(60,1){\framebox(5,3){$\FF_A$}}
\put(65,2.5){\line(1,0){5}}
\put(67,3){$\hat{A}$}
\put(70,1){\framebox(5,3){$\Phi_{\sim \hat{A}}$}}
\put(75,2.5){\line(1,0){5}}
\put(77,3){$\hat{A}$}
\put(80,1){\framebox(5,3){$\FF_A$}}
\put(85,2.5){\line(1,0){5}}
\put(87,3){$A$}
\put(92,2){$=$}
\put(95,2.5){\line(1,0){5}}
\put(97,3){$A$}

\end{picture}
\caption{Fourier transform identities.}
\label{Fig11}
\end{figure}

More simply, as a factor in a normal factor graph, the concatenation of $\FF_A$ with itself is equivalent up to scale to $\Phi_{\sim A}$, provided that the alphabet at each end of the chain is $A$;  see the top line of Figure \ref{Fig11}.  In this sense,
we may write $(\FF_A)^2 \propto \Phi_{\sim A}$, where the product indicates concatenation.  (Note that if the alphabet at each end of the chain were $\hat{A}$, then $(\FF_A)^2$ would be equivalent up to scale to $\Phi_{\sim \hat{A}}$, the sign inverter indicator function on $\hat{A}$.)

In the same sense, it is easy to see that $(\FF_A)^3 \propto \FF_A^*$, and $(\FF_A)^4 \propto I_A$.  In other words, $\FF_A$ is a fourth root of unity under concatenation, up to scale, and indeed behaves very much like $i = \sqrt{-1}$.  Similarly, $\FF_A^* \propto (\FF_A)^3$ is a fourth root of unity that is conjugate to $\FF_A$, while $\Phi_{\sim A} \propto (\FF_A)^2$ is a square root of unity and is real (equal to its conjugate).  

Figure \ref{Fig11} illustrates these relationships.  In this figure, edges are labelled simply by the alphabet of the associated variable rather than by the variable itself.

We remark that the relation $I_A \propto (\FF_A)^4 \propto \FF_A \Phi_{\sim \hat{A}} \FF_A$ is an instance of the Fourier transform relationship for orthogonal code indicator functions, since for any vector space $A$ and its dual $\hat{A}$, the orthogonal code to the sign inverter code $\CC_{\sim \hat{A}} =\{(\hat{a}, -\hat{a}): \hat{a} \in \hat{A}\}$, whose indicator function is $\Phi_{\sim \hat{A}}$, is the repetition code $\CC_{= A} = \{(a, a): a \in A\}$, whose indicator function is $I_A$.

Finally, note that in the special case where the characteristic of $\F$ is $p = 2$, we have that $\FF_A^* = \FF_A$ and $\Phi_{\sim A} = I_A$; \ie $\FF_A$ becomes a square root of unity, and the sign inverter relation  becomes the equality relation, $a' = a$.

\subsection{Normal factor graph duality theorem for linear codes}

We will now prove the normal graph duality theorem for linear codes by dualizing a normal factor graph. We continue to ignore scale factors.

Again, we start with a normal factor graph whose partition function is the indicator function $\Phi_\CC$ of a linear code $\CC\subseteq \A$, with half-edges representing external variables $\{A_k\}$, edges representing internal variables $\{S_j\}$, and vertices representing indicator functions $\{\Phi_{\CC_i}\}$.
%\pagebreak

\begin{figure}[h]
\setlength{\unitlength}{5pt}
\centering
\begin{picture}(36,14)(-2, 3)
\multiput(1,5)(12,0){3}{\line(1,0){6}}
\multiput(9.5,7.5)(12,0){2}{\line(0,1){3}}
\multiput(8,10.5)(12,0){2}{\framebox(3,3){$\FF$}}
\multiput(9.5,13.5)(12,0){2}{\line(0,1){3.5}}
\multiput(8,17)(12,0){2}{\line(1,0){3}}
\put(-2,5){\ldots}
\put(4,6){$S_k$}
\put(14,6){$S_{k+1}$}
\put(26,6){$S_{k+2}$}
\put(32,5){\ldots}
\put(10,8.5){$A_k$}
\put(10,14.5){$\hat{A}_k$}
\put(22,8.5){$A_{k+1}$}
\put(22,14.5){$\hat{A}_{k+1}$}
\put(7,2.5){\framebox(6,5){$\Phi_{\CC_k}$}}
\put(19,2.5){\framebox(6,5){$\Phi_{\CC_{k+1}}$}}
\end{picture}
\caption{Normal factor graph representing the orthogonal code to that of Figure \ref{Fig2}.}
\label{Fig12}
\end{figure}
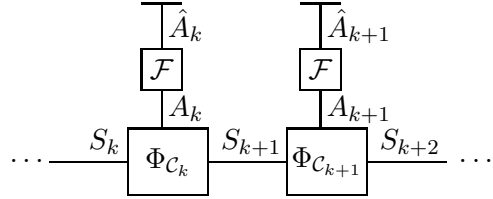

To obtain a normal factor graph whose partition function is the indicator function $\Phi_{\CC^\perp}$ of the orthogonal linear code $\CC^\perp \subseteq \hat{\A}$, we apply appropriate Fourier transforms $\FF$ to each of the external variables $A_k$, as in Figure 9.  For example, Figure \ref{Fig12} illustrates this transformation for two consecutive sections of a conventional state realization as in Figure \ref{Fig2}.

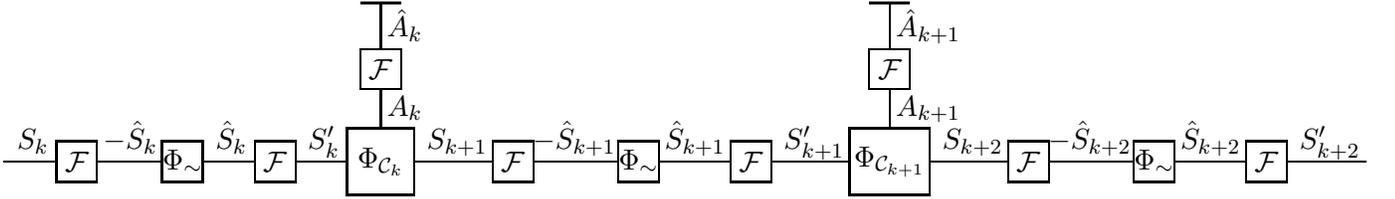
\begin{figure}[h]
\setlength{\unitlength}{5pt}
\centering
\begin{picture}(100,13)(5, 3)
\put(2,5){\line(1,0){4}}
\put(3,6){$S_k$}
\put(6,3.5){\framebox(3,3){$\FF$}}
\put(9,5){\line(1,0){5}}
\put(9.5,6){$-\hat{S}_k$}
\put(14,3.5){\framebox(3,3){$\Phi_\sim$}}
\put(17,5){\line(1,0){4}}
\put(18,6){$\hat{S}_k$}
\put(21,3.5){\framebox(3,3){$\FF$}}
\put(24,5){\line(1,0){4}}
\put(25,6){$S'_k$}
\put(28,2.5){\framebox(5,5){$\Phi_{\CC_k}$}}
\put(33,5){\line(1,0){6}}
\put(34,6){$S_{k+1}$}
\put(39,3.5){\framebox(3,3){$\FF$}}
\put(42,5){\line(1,0){6.5}}
\put(42,6){$-\hat{S}_{k+1}$}
\put(48.5,3.5){\framebox(3,3){$\Phi_\sim$}}
\put(51.5,5){\line(1,0){5.5}}
\put(52,6){$\hat{S}_{k+1}$}
\put(57,3.5){\framebox(3,3){$\FF$}}
\put(60,5){\line(1,0){6}}
\put(61,6){$S'_{k+1}$}
\put(66,2.5){\framebox(6,5){$\Phi_{\CC_{k+1}}$}}
\put(72,5){\line(1,0){6}}
\put(73,6){$S_{k+2}$}
\put(78,3.5){\framebox(3,3){$\FF$}}
\put(81,5){\line(1,0){6.5}}
\put(81,6){$-\hat{S}_{k+2}$}
\put(87.5,3.5){\framebox(3,3){$\Phi_\sim$}}
\put(90.5,5){\line(1,0){5.5}}
\put(91,6){$\hat{S}_{k+2}$}
\put(96,3.5){\framebox(3,3){$\FF$}}
\put(99,5){\line(1,0){6}}
\put(100,6){$S'_{k+2}$}
\put(30.5,7.5){\line(0,1){3}}
\put(31,8.5){$A_k$}
\put(29,10.5){\framebox(3,3){$\FF$}}
\put(30.5,13.5){\line(0,1){3.5}}
\put(31,14.5){$\hat{A}_k$}
\put(29,17){\line(1,0){3}}
\put(69,7.5){\line(0,1){3}}
\put(69.5,8.5){$A_{k+1}$}
\put(67.5,10.5){\framebox(3,3){$\FF$}}
\put(69,13.5){\line(0,1){3.5}}
\put(69.5,14.5){$\hat{A}_{k+1}$}
\put(67.5,17){\line(1,0){3}}
\end{picture}

\caption{Second step in dualizing Figure \ref{Fig2}.}
\label{Fig13}
\end{figure}

Next, using the concatenation relation $I \propto \FF \Phi_{\sim} \FF$ illustrated in Figure \ref{Fig11}, we replace each edge by the concatenation of an appropriate transform matrix $\FF$, sign inverter indicator function $\Phi_\sim$, and transform matrix $\FF$.  For example, Figure \ref{Fig13} illustrates this replacement for the normal factor graph of Figure \ref{Fig12}.  Here $S_k$ and $S_k'$ are two equal state variables, and dual state variables on opposite sides of sign inverters have been given opposite signs.

\begin{figure}[h]
\setlength{\unitlength}{5pt}
\centering
\begin{picture}(75,10)(5, 3)
\put(10,5){\ldots}
\put(15,5){\line(1,0){5}}
\put(15,6){$-\hat{S}_k$}
\put(20,3.5){\framebox(3,3){$\Phi_\sim$}}
\put(23,5){\line(1,0){5}}
\put(24,6){$\hat{S}_k$}
\put(28,2.5){\framebox(5,5){$\Phi_{\CC_k^\perp}$}}
\put(33,5){\line(1,0){6.5}}
\put(33,6){$-\hat{S}_{k+1}$}
\put(39.5,3.5){\framebox(3,3){$\Phi_\sim$}}
\put(42.5,5){\line(1,0){5.5}}
\put(43,6){$\hat{S}_{k+1}$}
\put(48,2.5){\framebox(6,5){$\Phi_{\CC_{k+1}^\perp}$}}
\put(54,5){\line(1,0){6.5}}
\put(54,6){$-\hat{S}_{k+2}$}
\put(60.5,3.5){\framebox(3,3){$\Phi_\sim$}}
\put(63.5,5){\line(1,0){5.5}}
\put(64,6){$\hat{S}_{k+2}$}
\put(70,5){\ldots}
\put(30.5,7.5){\line(0,1){3}}
\put(29,11.5){$\hat{A}_k$}
\put(29,10.5){\line(1,0){3}}
\put(51,7.5){\line(0,1){3}}
\put(49,11.5){$\hat{A}_{k+1}$}
\put(49.5,10.5){\line(1,0){3}}
\end{picture}

\caption{Final step in dualizing Figure 2.}
\label{Fig14}
\end{figure}
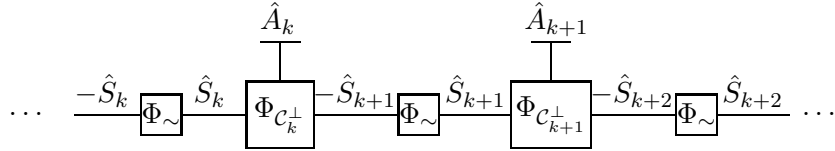

We then observe that every constraint code indicator function is now surrounded by Fourier transforms, as in Figure \ref{Fig9}, and therefore we may replace it by the indicator function of the dual constraint code, as shown in 
Figure \ref{Fig14}.  Note that the sign inverter indicator functions remain.

 In summary, we have proved:

\vspace{1ex}\pagebreak
\noindent
\textbf{Normal factor graph duality theorem for linear codes}. % \cite{MK05, AM10}.
Given a normal factor graph whose partition function is the indicator function $\Phi_\CC$ of a linear code $\CC$, comprising symbol alphabets $\{A_k$\} associated with half-edges, state spaces $\{S_j\}$ associated with ordinary edges, and constraint code indicator functions $\{\Phi_{\CC_i}\}$ associated with vertices, the dual normal factor graph is defined by replacing each symbol alphabet $A_k$ by its dual symbol alphabet $\hat{A}_k$, each state space $S_k$ by its dual state space $\hat{S}_k$, each indicator function $\Phi_{\CC_i}$ by the dual indicator function $\Phi_{\CC_i^\perp}$, and finally by placing a sign inverter indicator function $\Phi_{\sim}$ in the middle of every ordinary edge.  Then the partition function of the dual normal factor graph is equal to $\Phi_{\CC^\perp}$, up to scale. \qed
\vspace{1ex} 
%\pagebreak

%For example, Figure 10 shows the dual to the normal factor graph of a conventional state realization, as in Figure 2.
%%
%\begin{figure}[h]
%\setlength{\unitlength}{5pt}
%\centering
%\begin{picture}(70,12)(0, 1)
%\multiput(-1,5)(24,0){3}{\line(1,0){8}}
%\multiput(12,5)(24,0){3}{\line(1,0){8}}
%\multiput(9.5,7.5)(24,0){3}{\line(0,1){3}}
%\multiput(8,10.5)(24,0){3}{\line(1,0){3}}
%\put(-5,5){\ldots}
%\put(70,5){\ldots}
%\put(2,6){$\hat{S}_k$}
%\put(13,6){$\hat{S}_{k+1}$}
%\put(25,6){$\hat{S}_{k+1}$}
%\put(37,6){$\hat{S}_{k+2}$}
%\put(49,6){$\hat{S}_{k+2}$}
%\put(61,6){$\hat{S}_{k+3}$}
%\put(8,11.5){$\hat{A}_k$}
%\put(31,11.5){$\hat{A}_{k+1}$}
%\put(55,11.5){$\hat{A}_{k+2}$}
%\put(7,2.5){\framebox(5,5){$\Phi_{\CC_k^\perp}$}}
%\put(31,2.5){\framebox(5,5){$\Phi_{\CC_{k+1}^\perp}$}}
%\put(55,2.5){\framebox(5,5){$\Phi_{\CC_{k+2}^\perp}$}}
%\put(20,3.5){\framebox(3,3){$\Phi_{\sim}$}}
%\put(44,3.5){\framebox(3,3){$\Phi_{\sim}$}}
%\end{picture}

%\caption{Normal factor graph of dual of a conventional state realization.}
%\label{Fig10}
%\end{figure}

This theorem is equivalent to our original normal graph duality theorem \cite{F01}:  given a normal graph representing a linear code $\CC$, comprising symbol alphabets $\{A_k$\} associated with half-edges, state spaces $\{S_j\}$ associated with ordinary edges, and constraint codes $\{\CC_i\}$ associated with vertices, the dual normal graph is defined by replacing each symbol alphabet $A_k$ by its dual symbol alphabet $\hat{A}_k$, each state space $S_k$ by its dual state space $\hat{S}_k$, each constraint code $\CC_i$ by its orthogonal code $\CC_i^\perp$, and finally by placing a sign inverter in the middle of every ordinary edge.  Then the dual normal graph represents the orthogonal code $\CC^\perp$.

\begin{figure}[h]
\setlength{\unitlength}{5pt}
\centering
\begin{picture}(75,10)(5, 3)
\put(10,5){\ldots}
\put(15,5){\line(1,0){5}}
\put(15,6){$-\hat{S}_k$}
\put(20,3.5){\framebox(3,3){$\sim$}}
\put(23,5){\line(1,0){5}}
\put(24,6){$\hat{S}_k$}
\put(28,2.5){\framebox(5,5){${\CC_k^\perp}$}}
\put(33,5){\line(1,0){6.5}}
\put(33,6){$-\hat{S}_{k+1}$}
\put(39.5,3.5){\framebox(3,3){$\sim$}}
\put(42.5,5){\line(1,0){5.5}}
\put(43,6){$\hat{S}_{k+1}$}
\put(48,2.5){\framebox(6,5){${\CC_{k+1}^\perp}$}}
\put(54,5){\line(1,0){6.5}}
\put(54,6){$-\hat{S}_{k+2}$}
\put(60.5,3.5){\framebox(3,3){$\sim$}}
\put(63.5,5){\line(1,0){5.5}}
\put(64,6){$\hat{S}_{k+2}$}
\put(70,5){\ldots}
\put(30.5,7.5){\line(0,1){3}}
\put(29,11.5){$\hat{A}_k$}
\put(29,10.5){\line(1,0){3}}
\put(51,7.5){\line(0,1){3}}
\put(49,11.5){$\hat{A}_{k+1}$}
\put(49.5,10.5){\line(1,0){3}}
\end{picture}
\caption{Dual normal graph for a conventional state realization.}
\label{Fig15}
\end{figure}
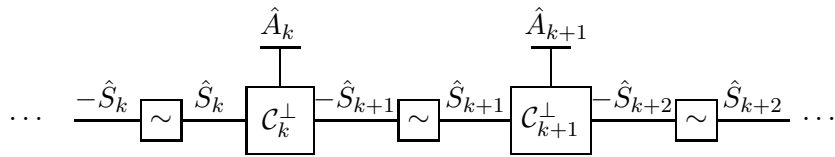

For example, Figure \ref{Fig15} shows the dual to the normal graph of a conventional state realization that was shown in Figure \ref{Fig1}.

The following two examples illustrate how to dualize binary and nonbinary convolutional codes by dualizing the local constraint codes $\CC_k$.  Example 3 involves a rate-1/2, four-state binary convolutional code that is often used in examples.  Example 4 involves the rate-2/3 ternary convolutional code that was the principal example of Gluesing-Luerssen and Schneider \cite{GLS08,GLS09}.

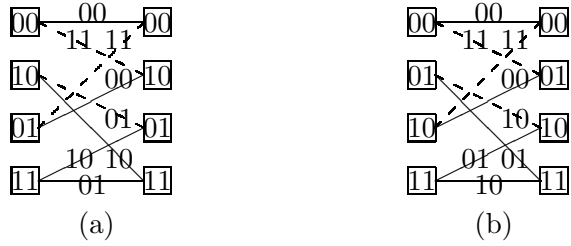
\begin{figure}[h]
\setlength{\unitlength}{5pt}
\centering
\begin{picture}(42,14)(0, -1)
\put(0,0){\framebox(2,2){$11$}}
\put(0,4){\framebox(2,2){$01$}}
\put(0,8){\framebox(2,2){$10$}}
\put(0,12){\framebox(2,2){$00$}}
\put(10,0){\framebox(2,2){$11$}}
\put(10,4){\framebox(2,2){$01$}}
\put(10,8){\framebox(2,2){$10$}}
\put(10,12){\framebox(2,2){$00$}}
\put(2,1){\line(1,0){8}}
\put(2,1){\line(2,1){8}}
\put(2,5){\line(2,1){8}}
\put(2,13){\line(1,0){8}}
\put(2,9){\line(1,-1){8}}
\dashline{0.75}(2,13)(10,9)
\dashline{0.75}(2,9)(10,5)
\dashline{0.75}(2,5)(10,13)
\put(5,0){$01$}
\put(4,2){$10$}
\put(7,5){$01$}
\put(7,2){$10$}
\put(5,13){$00$}
\put(4,11){$11$}
\put(7,8){$00$}
\put(7,11){$11$}
\put(5,-3){(a)}

%\put(20,6){$\Leftrightarrow$}

\put(30,0){\framebox(2,2){$11$}}
\put(30,4){\framebox(2,2){$10$}}
\put(30,8){\framebox(2,2){$01$}}
\put(30,12){\framebox(2,2){$00$}}
\put(40,0){\framebox(2,2){$11$}}
\put(40,4){\framebox(2,2){$10$}}
\put(40,8){\framebox(2,2){$01$}}
\put(40,12){\framebox(2,2){$00$}}
\put(32,1){\line(1,0){8}}
\put(32,1){\line(2,1){8}}
\put(32,5){\line(2,1){8}}
\put(32,13){\line(1,0){8}}
\put(32,9){\line(1,-1){8}}
\dashline{0.75}(32,13)(40,9)
\dashline{0.75}(32,9)(40,5)
\dashline{0.75}(32,5)(40,13)
\put(35,0){$10$}
\put(34,2){$01$}
\put(37,5){$10$}
\put(37,2){$01$}
\put(35,13){$00$}
\put(34,11){$11$}
\put(37,8){$00$}
\put(37,11){$11$}
\put(35,-3){(b)}

\end{picture}

\caption{Trellis sections of (a) rate-1/2  binary convolutional code; (b) orthogonal code.}
\label{Fig16}
\end{figure}

\vspace{1ex}
\noindent
\textbf{Example 3} (binary linear convolutional code).  Consider the rate-1/2 binary linear time-invariant convolutional code $\CC$ generated by the degree-2 generators $(1 + D^2, 1 + D + D^2)$, in standard $D$-transform notation.  In other words, $\CC$ is the set of all output sequences of the single-input, two-output linear time-invariant system over $\F_2$ whose impulse response is $(11, 01, 11, 00, \ldots)$. This system has a conventional state realization as in Figure \ref{Fig1} in which each symbol alphabet $A_k$ may be taken as $(\F_2)^2$, each state space $S_k$ may also be taken as $(\F_2)^2$, and each constraint code $\CC_k$ is the $(6, 3)$ binary linear block code generated by the three transitions
$$
\begin{array}{ccc}
00 & 11 & 10 \\
10 & 01 & 01 \\
01 & 11 & 00
\end{array}
$$
which represent the three nontrivial (state, symbol, next state) transitions in the impulse response of the system.  (Note that only the output symbols appear in $\CC$;  the input symbols that would appear in an input-state-output realization are here regarded as internal variables, and do not appear explicitly.)   The eight codewords of $\CC_k$ are the eight possible transitions of the system, which are shown as a ``trellis section" in Figure \ref{Fig16}(a), with the three generating transitions dashed.

%

%\noindent
The orthogonal code $\CC_k^\perp$ may easily be seen to be  the $(6, 3)$ binary linear block code generated by the three transitions
$$
\begin{array}{ccc}
00 & 11 & 01 \\
01 & 10 & 10 \\
10 & 11 & 00
\end{array}
$$
which represent the three nontrivial (state, symbol, next state) transitions in the impulse response of a system with impulse response $(11, 10, 11, 00, \ldots)$, or $(1 + D + D^2, 1 + D^2)$ in $D$-transform notation.  This is indeed the generator of the orthogonal convolutional code $\CC^\perp$ under the symbolwise definition of the inner product that we are using here.  (For the more usual sequencewise definition of the inner product, we need to take the time reversal of $\CC^\perp$,\footnote{The symbolwise inner product of two sequences $\ab, \bb \in \A$ is $\sum_{k} a_k b_k$, and that of $\ab$ and a shift of $\bb$ by $j$ time units is $\sum_{k} a_k b_{k-j}$.  The product of the corresponding $D$-transforms $a(D) = \sum_k a_kD^k$ and $b(D^{-1}) = \sum_k b_kD^{-k}$ is $\sum_j (\sum_{k} a_k b_{k-j})D^j$, so $\ab$ is orthogonal to all shifts of $\bb$ if and only if $a(D)b(D^{-1}) = 0$, or equivalently if and only if the sequencewise inner product $a(D)\tilde{b}(D)$ is zero, where $\tilde{b}(D)$ is the $D$-transform of the time-reversed sequence $\tilde{\bb} = \{b_{-k}: k \in \I_\A\}$. Thus $\D$ is the orthogonal code to a linear time-invariant code $\CC$ under the symbolwise inner product if and only if the time-reversed code $\tilde{\D}$ is the orthogonal code to $\CC$ under the sequencewise inner product.} which in this case happens to be the code generated by $(1 + D + D^2, 1 + D^2)$ again.)  The eight codewords of $\CC_k^\perp$ are the eight possible transitions of the orthogonal system, which are again shown as a ``trellis section" in Figure \ref{Fig16}(b), with the three generating transitions dashed.  \qed %\vspace{1ex}

\vspace{1ex}
\noindent
\textbf{Example 4} (ternary linear convolutional code; \emph{cf.\ }\cite{GLS08, GLS09}).  Consider the rate-2/3 linear time-invariant convolutional code $\CC$ over $\F_3$ with $g_1(D) = (1 + D^2, 2 + D, 0)$ and $g_2(D) = (1, 0, 2)$.  In other words, $\CC$ is the set of all output sequences of the two-input, three-output linear time-invariant system over $\F_3$ whose impulse responses are $(120, 010, 100, 000, \ldots)$ and $(102, 000, \ldots).$  This system has a conventional nine-state realization as in Figure \ref{Fig1} in which each symbol alphabet $A_k$ may be taken as $(\F_3)^3$, each state space $S_k$ may be taken as $(\F_3)^2$, and each constraint code $\CC_k$ is the $(7, 4)$ ternary linear block code generated by the four generators  
$$
\begin{array}{ccc}
00 & 120 & 10 \\
10 & 010 & 01 \\
01 & 100 & 00 \\
00 & 102 & 00
\end{array}
$$
which represent the four nontrivial ($s_k, a_k, s_{k+1}$) transitions in the two impulse responses of the system.  The orthogonal code $\CC_k^\perp$ is  the $(7, 3)$ ternary linear block code generated by the three generators
$$
\begin{array}{ccc}
00 & 010 & 12 \\
21 & 202 & 11 \\
22 & 111 & 00
\end{array}
$$
which represent the three nontrivial ($\hat{s}_k, \hat{a}_k, -\hat{s}_{k+1}$) transitions in the impulse response of a conventional state realization of a single-input, three-output linear system over $\F_3$, with sign inverters, whose impulse response is $(010, 202, 111, 000, \ldots)$, or $(2D + D^2, 1 + D^2, 2D + D^2)$ in $D$-transform notation.  (Note the unconventional basis of the dual state space, and the effects of the sign inversions.)  This is indeed the generator of the orthogonal convolutional code $\CC^\perp$ under our symbolwise definition of the inner product.  (For the more usual sequencewise definition of the inner product, we need to take the time reversal of $\CC^\perp$, which in this case is the code generated by $(1 + 2D, 1 + D^2, 1 + 2D)$.)\qed

\subsection{General normal factor graph duality theorem}

Finally, we note that although we have been focussing on normal factor graphs whose factors and partition functions are indicator functions of linear codes, the proof of the normal factor graph duality theorem in the previous subsection does not depend upon this restriction.  Thus we have actually proved a much more general theorem:

\vspace{1ex}
\noindent
\textbf{General normal factor graph duality theorem}. 
Given an arbitrary normal factor graph with partition function $Z(\ab)$, up to scale, comprising external variables $\{A_k$\} associated with half-edges, internal variables $\{S_j\}$ associated with ordinary edges, and factors $\{f_i\}$ associated with vertices, the dual normal factor graph is defined by replacing each external variable $A_k$ by its dual variable $\hat{A}_k$, each internal variable $S_j$ by its dual variable $\hat{S}_j$, each factor $f_i$ by its Fourier transform $\hat{f}_i$, and finally by placing a sign inverter indicator function $\Phi_{\sim}$ in the middle of every ordinary edge.  Then the partition function of the dual normal factor graph is the Fourier transform $\hat{Z}(\hat{\ab})$ of $Z(\ab)$, up to scale.
(Note: \cite{AM10} shows that the scale factor is $\prod_j |\SSS_j|$.)  \qed \vspace{1ex}

Mao and Kschischang \cite{MK05} have proved a closely related duality theorem.  In their development, a (multiplicative) factor graph represents a product of factors, rather than a sum of products.  They show that the dual to a multiplicative factor graph is a \emph{convolutional} factor graph, which represents the convolution of its factors.  If the elements of a factor graph are dualized as above (but without the sign inverters), then a factor graph and its dual represent a Fourier transform pair, up to scale.  Using this duality result and the fact that the indicator functions of a linear code and its dual are a Fourier transform pair, they derive a factor graph duality theorem for linear codes.  They then specialize to normal factor graphs, and show how to recover our original normal graph duality theorem for linear codes.  A similar derivation would yield the general normal factor graph duality theorem above.  The advantage of our approach is that by focussing on normal factor graphs and their partition functions from the beginning, we do not need to introduce convolutional factor graphs (which may however prove to have their uses). 

Al-Bashabsheh and Mao \cite{AM10} have also proved this theorem by their methods, independently and at about the same time.   Loeliger \emph{et al.\ }\cite[Appendix III]{L07} have proved an analogous theorem for continuous alphabets.

\pagebreak
\section{MacWilliams identities}

Given these duality results, we can obtain various MacWilliams identities, using similar derivations.

\subsection{MacWilliams identities for exact weight generating functions}

We will first consider what MacWilliams and Sloane \cite{MS77} call \emph{exact} weight generating functions, where every symbol alphabet $A_k$ is given its own set of indeterminates $\xb_k = \{x_k(a_k):  a_k \in A_k\}$.  The dual set of indeterminates $\Xb_k = \{X_k(\hat{a}_k):  \hat{a}_k \in \hat{A}_k\}$ is then given by the transform $\Xb_k = \FF_{A_k} \xb_k$.

The \emph{exact weight generating function} (EWGF) of $\CC$  is the multivariate polynomial
$$
g^E_\CC(\xb) = \sum_{\ab \in \CC} \xb(\ab) = \sum_{\ab \in \A} \Phi_\CC(\ab) \xb(\ab),
$$
where $\xb(\ab) = \prod_{k \in \I_\A} x_k(a_k)$.
%Thus for any $\ab \in \A$, the coefficient of $\xb(\ab)$ in $g^E_\CC(\xb)$ is $\Phi_\CC(\ab)$.

Similarly, the EWGF of $\CC^\perp$  is the multivariate polynomial
$$
g^E_{\CC^\perp}(\Xb) = \sum_{\hat{\ab} \in \hat{\A}} \Phi_{\CC^\perp}(\hat{\ab}) \Xb(\hat{\ab}).
$$
By the Poisson summation formula, these dual EWGFs are equal, up to scale;  see Figure \ref{Fig17}.  (Again we note that since both $\Phi_\CC$ and $\Phi_{\CC^\perp}$ are real-valued, the Fourier transform relation holds with the kernel $\FF_\A^*$ as well as with $\FF_\A$.)  Explicitly, we have the following MacWilliams identity showing how the EWGF of $\CC^\perp$ may be computed from that of $\CC$:
$$
g^E_{\CC^\perp}(\Xb) \propto g^E_{\CC}(\FF_{\A}^{*}\Xb)
$$

\begin{figure}[h]
\setlength{\unitlength}{5pt}
\centering
\begin{picture}(70,3)(8, 7)
\put(10,6){\framebox(5,3){$\Phi_\CC$}}
\put(15,7.5){\line(1,0){5}}
\put(17,8){$\A$}
\put(20,6){\framebox(5,3){$\xb$}}
\put(27,7){$=$}
\put(30,6){\framebox(5,3){$\Phi_\CC$}}
\put(35,7.5){\line(1,0){5}}
\put(37,8){$\A$}
\put(40,6){\framebox(5,3){$\FF_{\A}^*$}}
\put(45,7.5){\line(1,0){5}}
\put(47,8){$\hat{\A}$}
\put(50,6){\framebox(5,3){$\Xb$}}
\put(57,7){$=$}
\put(60,6){\framebox(5,3){$\Phi_{\CC^\perp}$}}
\put(65,7.5){\line(1,0){5}}
\put(67,8){$\hat{\A}$}
\put(70,6){\framebox(5,3){$\Xb$}}
\end{picture}
\caption{Duality of exact weight generating functions.}
\label{Fig17}
\end{figure}

Now again suppose that we have a normal factor graph whose partition function is $\Phi_\CC$ (up to scale), defined by a set $\Ab = \{A_k\}$ of symbol alphabets, a set $\Sb = \{S_j\}$ of state spaces, and a set $\{\Phi_{\CC_i}\}$ of constraint code indicator functions, each constraint code $\CC_i$ constraining subsets $\Ab_i \subseteq \Ab, \Sb_i \subseteq \Sb$ of the symbol and state alphabets, respectively.  Then we may obtain a normal factor graph whose partition function is the global EWGF $g^E_\CC(\xb)$ by connecting each half-edge representing $A_k$ to a corresponding indeterminate function $x_k:  a_k \mapsto x_k(a_k)$, as in Figure \ref{Fig3}.  The \emph{local exact weight generating function} $g^E_{\CC_i}(\xb_i, \sb_i)$ is then obtained by summing over $\Ab_i$ for each $\CC_i$ to obtain
$$
g^E_{\CC_i}(\xb_i, \sb_i) = \sum_{\ab_i \in \Ab_i} \Phi_{\CC_i}(\ab_i, \sb_i) \xb_i(\ab_i).
$$
The global EWGF is then the resulting global partition function, up to scale:
$$
g^E_\CC(\xb) = \sum_{\sb \in \SSS} \prod_{i \in \I_\CC} g^E_{\CC_i}(\xb_i, \sb_i).
$$

In the case of a conventional state realization, the local EWGF is a matrix indexed by $S_k \times S_{k+1}$,
$$\Lambda_k^E(x_k) = \{g^E_{\CC_k}(x_k, s_k, s_{k+1}):  s_k \in S_k, s_{k+1} \in S_{k+1}\},$$
 which we will call the \emph{exact weight adjacency matrix} (EWAM), and the global EWGF is the matrix product of these EWAMs, as was shown for HWGFs and HWAMs in Figures \ref{Fig4} and \ref{Fig5}.

In general, if the graph of a code is a tree (\ie cycle-free), then since a global generating function of the code is a sum of products of local factors, it may be computed by a straightforward application of the generic sum-product algorithm \cite{F01}, in which the ``messages" are generating functions of subtrees.
Even if the graph is not a tree, the global generating function is still the sum of the products of the local generating functions over all $\sb \in \SSS$.  Consequently one method of computing it is to cut just enough state edges so that the graph becomes a tree (a minimal spanning tree), compute the global generating function for this tree using the sum-product algorithm, and then sum over the remaining state variables.
%\pagebreak

\vspace{1ex}
\noindent
\textbf{Example 5} (tail-biting trellis).  
A tail-biting trellis consists of a chain of $N$ trellis sections, with the further constraint that the final state $s_N$ is equal to the initial state $s_0$.  To compute a global generating function of a tail-biting trellis, we may thus first form the matrix product of the $N$ constituent local WAMs to obtain a global WAM, indexed by $(s_0, s_N)$;  we then sum over all elements of this WAM for which $s_0 = s_N$;  \ie we take the \emph{trace} of the global WAM. \qed \vspace{1ex}
 
\begin{figure}[h]
\setlength{\unitlength}{5pt}
\centering
\begin{picture}(78,14)(8, 6)
\put(10,6){\framebox(5,13){$\Phi_{\CC_i}$}}
\put(15,17.5){\line(1,0){5}}
\put(16,18.5){$\Ab_i$}
\put(20,16){\framebox(5,3){$\xb_i$}}
\put(15,12.5){\line(1,0){10}}
\put(16,13.5){$\Sb_i^+$}
\put(25,11){\line(0,1){3}}
\put(15,7.5){\line(1,0){10}}
\put(16,8.5){$\Sb_i^-$}
\put(25,6){\line(0,1){3}}
\put(16,4){(a)}
\put(27,12){$\Rightarrow$}
\put(30,6){\framebox(5,13){$\Phi_{\CC_i}$}}
\put(35,17.5){\line(1,0){5}}
\put(36,18.5){$\Ab_i$}
\put(40,16){\framebox(5,3){$\FF^*$}}
\put(45,17.5){\line(1,0){5}}
\put(46,18.5){$\hat{\Ab}_i$}
\put(50,16){\framebox(5,3){$\Xb_i$}}
\put(35,12.5){\line(1,0){5}}
\put(36,13.5){$\Sb_i^+$}
\put(40,11){\framebox(5,3){$\FF^*$}}
\put(45,12.5){\line(1,0){14}}
\put(46,13.5){$\hat{\Sb}_i^+$}
\put(59,11){\line(0,1){3}}
\put(35,7.5){\line(1,0){5}}
\put(36,8.5){$\Sb_i^-$}
\put(40,6){\framebox(5,3){$\FF^*$}}
\put(45,7.5){\line(1,0){5}}
\put(46,8.5){$\hat{\Sb}_i^-$}
\put(50,6){\framebox(3,3){$\Phi_\sim$}}
\put(53,7.5){\line(1,0){6}}
\put(54,8.5){$-\hat{\Sb}_i^-$}
\put(59,6){\line(0,1){3}}
\put(36,4){(b)}
\put(62,12){$=$}
\put(65,6){\framebox(5,13){$\Phi_{\CC_i^\perp}$}}
\put(70,17.5){\line(1,0){5}}
\put(71,18.5){$\hat{\Ab}_i$}
\put(75,16){\framebox(5,3){$\Xb_i$}}
\put(70,12.5){\line(1,0){14}}
\put(71,13.5){$\hat{\Sb}_i^+$}
\put(84,11){\line(0,1){3}}
\put(70,7.5){\line(1,0){5}}
\put(71,8.5){$\hat{\Sb}_i^-$}
\put(75,6){\framebox(3,3){$\Phi_\sim$}}
\put(78,7.5){\line(1,0){6}}
\put(79,8.5){$-\hat{\Sb}_i^-$}
\put(84,6){\line(0,1){3}}
\put(71,4){(c)}
\end{picture}
\caption{Dualizing local exact weight generating functions.}
\label{Fig18}
\end{figure}
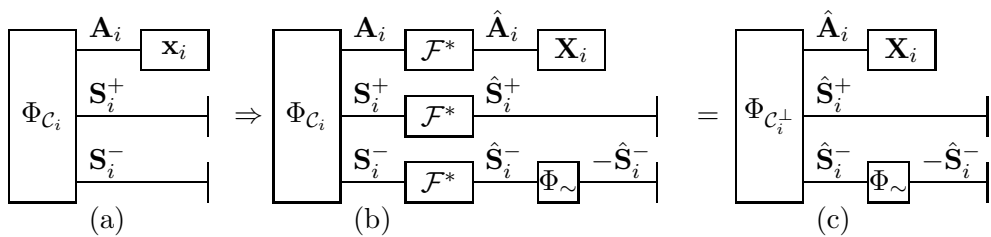

To obtain a MacWilliams identity for local EWGFs, we proceed as follows.  Each local EWGF $g^E_{\CC_i}(\xb_i, \sb_i)$ is the partition function of the local graph fragment shown in Figure \ref{Fig18}(a), comprising a local constraint code indicator function $\Phi_{\CC_i}$ attached via an edge labelled by the local symbol alphabets $\Ab_i$ to an exact weight generator function $\xb_i$.  Here we partition the local state spaces $\Sb_i$ into two subsets $\Sb_i^+$ and $\Sb_i^-$, such that globally every state space $S_j$ appears once in a plus subset and once in a minus subset (corresponding to the sign inversion in the dual normal factor graph).  Thus the local EWGF will now be written as $g^E_{\CC_i}(\xb_i, \sb_i^+, \sb^-_i)$.

Proceeding again along the lines of the derivation of the Poisson summation formula (see Figure \ref{Fig10}), we now replace the function $\xb_i$ by the concatenation of $\Xb_i$ and an appropriate inverse Fourier transform $\FF^*$;  we adjoin appropriate inverse Fourier transforms $\FF^*$ to each $S_j \in \Sb_i^+$; and we adjoin appropriate inverse Fourier transforms $\FF^*$ and sign inverter indicator functions $\Phi_\sim$ to each $S_j \in \Sb_i^-$, as shown in Figure \ref{Fig18}(b).  Because the concatenation of $\FF^*$, $\Phi_\sim$ and $\FF^*$ is the identity (see Figure \ref{Fig11}), and because globally every state space $S_j$ appears once in a plus subset and once in a minus subset, this will leave the global EWGF unchanged.

Finally, noticing that each code constraint indicator function $\Phi_{\CC_i}$ is now surrounded by (inverse) Fourier transforms, we may replace the whole ensemble by the orthogonal code indicator function $\Phi_{\CC_i^\perp}$, as shown in Figure \ref{Fig18}(c).  Globally, it is evident that we now have a factor graph whose partition function is $g^E_{\CC^\perp}(\Xb)$, comprising the dual normal factor graph for $\Phi_{\CC^\perp}$, with each half-edge representing $\hat{\Ab}_i$ connected to a corresponding indeterminate function $\Xb_i$.

Now we have the following MacWilliams identity, showing how the local EWGF of $\CC_i^\perp$ may be computed from that of $\CC_i$:
$$
g^E_{\CC_i^\perp}(\Xb_i, \hat{\Sb}_i^+, \hat{\Sb}_i^-) \propto g^E_{\CC_i}(\FF_{\Ab_i}^{*}\Xb_i, \FF_{\Sb_i^+}^*\Sb_i^+, \FF_{\Sb_i^-}(-\Sb_i^-)),
$$
where $\FF_{\Sb_i^-}^*\Sb_i^- \propto  \FF_{\Sb_i^-}(-\Sb_i^-)$ since the variables in $\Sb_i^+$ are the same as those in some other $-\Sb_i^-$.

In particular, with a conventional state realization with constraint codes $\CC_k$, with the convention that  $\hat{S}_k$ is involved in $\CC_k^\perp$ with a positive sign and $\hat{S}_{k+1}$ is involved in $\CC_k^\perp$ with a negative sign, we have the following MacWilliams identity, which shows how the EWAM $\hat{\Lambda}_k(\Xb_{k})$ of $\CC_k^\perp$ may be obtained from the EWAM $\Lambda_k(\xb_{k})$ of $\CC_k$:
$$
\hat{\Lambda}^E_k(\Xb_{k}) \propto (\FF_{S_k}^{*})^T \Lambda^E_k(\FF_{A_k}^{*}\Xb_{k}) ( \FF_{S_{k+1}})^T.
$$
(Here the transposes are needed to conform to standard matrix conventions, even though $\FF = \FF^T$.)

%\pagebreak
\vspace{1ex}
\noindent
\textbf{Example 3} (cont.).  Continuing with the rate-$\half$ binary convolutional code of Example 3, each constraint code $\CC_k$ has the exact weight adjacency matrix below, where we write $x_{00}, x_{01}, x_{10}, x_{11}$ for $x_{k}(00), x_{k}(01), x_{k}(10), x_{k}(11)$, respectively.
$$
\Lambda^E_k(\xb_{k}) \quad  = \quad
\begin{array}{c|c|c|c|c|}
s_k/s_{k+1} & 00 & 10 & 01 & 11 \\
\hline
00 & x_{00} & x_{11} & 0 & 0 \\
\hline
10 & 0 & 0 & x_{01} & x_{10} \\
\hline
01 & x_{11} & x_{00} & 0 & 0 \\
\hline
11 & 0 & 0  &x_{10} & x_{01} \\
\hline
\end{array}
$$
Given this EWAM for $\CC_k$, the EWAM  $\hat{\Lambda}^E_k(\Xb_{k})$ of the orthogonal constraint code $\CC_k^\perp$ is given by the matrix equation below, where we have substituted the dual indeterminates $X_{00}, X_{10}, X_{01}$ and $X_{11}$ for $x_{00} + x_{10} + x_{01} + x_{11}, x_{00} - x_{10} + x_{01} - x_{11}, x_{00} + x_{10} - x_{01} - x_{11}$ and $x_{00} - x_{10} - x_{01} + x_{11}$, respectively.  (We have also inserted the correct scale factor.)
$$
\half
 \left[ \begin{array}{crrr}
1 & 1 & 1 & 1 \\
1 & -1 & 1 & -1 \\
1 & 1 & -1 & -1 \\
1 & -1 & -1 & 1 \\
\end{array}  \right]
 \left[ \begin{array}{cccc}
x_{00} & x_{11} & 0 & 0 \\
0 & 0 & x_{01} & x_{10} \\
x_{11} & x_{00} & 0 & 0 \\
0 & 0  & x_{10} & x_{01} \\
\end{array}  \right]
 \left[ \begin{array}{crrr}
1 & 1 & 1 & 1 \\
1 & -1 & 1 & -1 \\
1 & 1 & -1 & -1 \\
1 & -1 & -1 & 1 \\
\end{array}  \right]
=
 \left[ \begin{array}{cccc}
X_{00} & 0 & X_{11} & 0 \\
X_{11} & 0 & X_{00} & 0 \\
0 & X_{10} & 0 & X_{01} \\
0 & X_{01} & 0 & X_{10} \\
\end{array}  \right]
$$
From Figure \ref{Fig16}(b), we see that this matrix is indeed the EWAM of the constraint code $\CC_k^\perp$. \qed \vspace{1ex}

Of course it is no surprise that a dual constraint code $\CC_i^\perp$ is determined by $\CC_i$, or \emph{vice versa};  what the MacWilliams identities give us is a way of \emph{calculating} one from the other, by taking transforms.

\subsection{MacWilliams identities for other weight generating functions}

Commonly each symbol alphabet $A_k$ is equal to $\F^{n_k}$, the set of $n_k$-tuples over the ground field $\F$;  \ie $a_k = \{a_{k\ell} \in \F: 1 \le \ell \le n_k\}$.  In this case a transform over $A_k$ may be expressed by $n_k$ transforms over $\F$ applied to the $n_k$ components $a_{k\ell}$ of $A_k$.  

We may then be interested in the \emph{complete weight generating function} (CWGF)
of $\CC$, defined as
$$
g_{\CC}^C(\xb) = \sum_{\ab \in \A} \Phi_{\CC}(\ab) \left( \prod_{k \in \I_\A} \prod_{\ell = 1}^{n_k} x(a_{k\ell}) \right),
$$
where $\xb = \{x(a): a \in \F\}$ is a set of indeterminates defined on $\F$.  In other words, the CWGF is obtained by substituting the product $\prod_{\ell = 1}^{n_k} x(a_{k\ell})$ for $x_{k}(a_k)$ in the EWGF.

Similarly, the \emph{local complete weight generating function} $g^C_{\CC_i}(\xb_i, \sb_i)$ is
$$
g^C_{\CC_i}(\xb, \sb_i) = \sum_{\ab_i \in \Ab_i} \Phi_{\CC_i}(\ab_i, \sb_i) \left( \prod_{k \in \I_{\Ab_i}} \prod_{\ell = 1}^{n_k} x(a_{k\ell}) \right),
$$
where $\I_{\Ab_i} \subseteq \I_\A$ denotes the subset of indices of symbol alphabets $A_k$ that are involved in $\CC_i$.

By a development parallel to that above, we have the following MacWilliams identity, which shows how the local CWGF of $\CC_i^\perp$ may be computed from that of $\CC_i$:
$$
g^C_{\CC_i^\perp}(\Xb, \hat{\Sb}_i^+, \hat{\Sb}_i^-) \propto g^C_{\CC_i}(\FF_{\F}^{*}\Xb, \FF_{\Sb_i^+}^*\Sb_i^+, \FF_{\Sb_i^-}(-\Sb_i^-)),
$$
where $\Xb = \FF_{\F} \xb$ is the dual set of indeterminates to the set $\xb$, with $\FF_{\F}$ the transform matrix on $\F$.  
%\pagebreak

In particular, for conventional state realizations,  the following MacWilliams identity shows how the complete weight adjacency matrix (CWAM) $\hat{\Lambda}_k^C(\Xb)$ of $\CC_k^\perp$ may be obtained from the CWAM $\Lambda_k^C(\xb)$ of $\CC_k$:
$$
\hat{\Lambda}^C_k(\Xb_{k}) \propto (\FF_{S_k}^{*})^T \Lambda^C_k(\FF_{\F}^{*}\Xb)  (\FF_{S_{k+1}})^T.
$$ 

\vspace{1ex}
\noindent
\textbf{Example 3} (cont.).  For the rate-$\half$ binary convolutional code of Example 3, each constraint code $\CC_k$ has the CWAM
$$
\Lambda_k^C(\xb) \quad  = \quad
\begin{array}{c|c|c|c|c|}
s_k/s_{k+1} & 00 & 10 & 01 & 11 \\
\hline
00 & x_{0}^2 & x_{1}^2 & 0 & 0 \\
\hline
10 & 0 & 0 & x_{0}x_{1} & x_{0}x_{1} \\
\hline
01 & x_{1}^2 & x_{0}^2 & 0 & 0 \\
\hline
11 & 0 & 0  & x_{0}x_{1} & x_{0}x_{1} \\
\hline
\end{array}
$$
where we have written $x_{0}$ and $x_{1}$ instead of $x(0)$ and $x(1)$, respectively.  The CWAM  $\hat{\Lambda}_k^C(\Xb)$ of the orthogonal constraint code $\CC_k^\perp$ is given by the matrix equation below, where we have substituted the dual indeterminates $X_{0}$ and $X_{1}$ for $x_{0} + x_{1}$ and $x_{0} - x_{1}$, respectively. 
$$
\half
 \left[ \begin{array}{crrr}
1 & 1 & 1 & 1 \\
1 & -1 & 1 & -1 \\
1 & 1 & -1 & -1 \\
1 & -1 & -1 & 1 \\
\end{array}  \right]
 \left[ \begin{array}{cccc}
x_{0}^2 & x_{1}^2 & 0 & 0 \\
0 & 0 &x_{0}x_{1} & x_{0}x_{1} \\
x_{1}^2 & x_{0}^2 & 0 & 0 \\
0 & 0  & x_{0}x_{1} & x_{0}x_{1} \\
\end{array}  \right]
 \left[ \begin{array}{crrr}
1 & 1 & 1 & 1 \\
1 & -1 & 1 & -1 \\
1 & 1 & -1 & -1 \\
1 & -1 & -1 & 1 \\
\end{array}  \right]
=
 \left[ \begin{array}{cccc}
X_{0}^2 & 0 & X_{1}^2 & 0 \\
X_{1}^2 & 0 & X_{0}^2 & 0 \\
0 & X_{0}X_{1} & 0 & X_{0}X_{1} \\
0 & X_{0}X_{1} & 0 & X_{0}X_{1} \\
\end{array}  \right]
$$
We see from Figure \ref{Fig16}(b) that this matrix is indeed the CWAM of  $\CC_k^\perp$. \qed \vspace{1ex}

Here the point is that even though the CWGF of $\CC_i$ does not fully determine $\CC_i$, it does determine the CWGF of $\CC_i^\perp$, and \emph{vice versa}.

%\pagebreak
The \emph{Hamming weight generating function} (HWGF) $g_{\CC}^H(x)$ of a linear code $\CC$ may be obtained by substituting $1$ for $x(0)$ and $x$ for each $x(a), a \neq 0$, in $g_{\CC}^C(\xb)$, and similarly for the \emph{Hamming weight adjacency matrix} (HWAM) $\Lambda_k^H(x)$ of a constraint code $\CC_k$ in a conventional state realization.  Thus each element of $\Lambda_k^H(x)$ becomes a polynomial of degree $n_k$ or less in the single indeterminate $x$.  The dual indeterminates become $X(0) = 1 + (|\F| - 1)x$ and $X(\hat{a}) = 1 - x$ for $\hat{a} \neq 0$, which scale to $1$ and $X = (1-x)/(1 + (|\F|-1)x)$, respectively.  Substituting in the above MacWilliams identities for CWGFs or CWAMs, we obtain MacWilliams identities for HWGFs or HWAMs.  This yields the main result of \cite{GLS08, GLS09}.\footnote{The MacWilliams identity of \cite{GLS08, GLS09} is stated in terms of the HWAM for a minimal realization of a linear time-invariant convolutional code $\CC$ in controller canonical form, and the HWAM of \emph{some} minimal encoder for the orthogonal code $\CC^\perp$.  Our results apply to the CWAM or HWAM of any state realization, and the CWAM or HWAM of its dual realization, because in our development, by constraint code duality, the basis of the dual state space representation is fixed as soon as the basis of the primal state space is fixed.}   

\vspace{1ex}
\noindent
\textbf{Example 3 (cont.)}.  For the rate-1/2 binary convolutional code $\CC$ of Example 3, each constraint code $\CC_k$ has the HWAM
$$
\Lambda^H(x) \quad  = \quad
\begin{array}{c|c|c|c|c|}
s/s' & 00 & 10 & 01 & 11 \\
\hline
00 & 1 & x^2 & 0 & 0 \\
\hline
10 & 0 & 0 & x & x \\
\hline
01 & x^2 & 1 & 0 & 0 \\
\hline
11 & 0 & 0  & x & x \\
\hline
\end{array}
$$
For the orthogonal code $\CC^\perp$, each constraint code $\CC^\perp_k$ has the HWAM
$$
\hat{\Lambda}^H(X) \quad  = \quad
\begin{array}{c|c|c|c|c|}
s/s' & 00 & 10 & 01 & 11 \\
\hline
00 & 1 & 0 & X^2 & 0 \\
\hline
10 & X^2 & 0 & 1 & 0 \\
\hline
01 & 0 & X & 0 & X \\
\hline
11 & 0 & X  & 0 & X \\
\hline
\end{array}
$$
The reader may verify that $\Lambda^H(x)$ and $\hat{\Lambda}^H(X)$ satisfy a MacWilliams identity with $X = \frac{1-x}{1+x}$.  [Hint:  it may be easier to start with CWAMs.]
Note that here $\hat{\Lambda}^H(x)$ happens to be the transpose of $\Lambda^H(x)$. \qed
\vspace{1ex}

%\vspace{1ex}
\noindent
\textbf{Example 4 (cont.)}.  For a worked-out example of  the HWAM  $\hat{\Lambda}_k^H(X)$ of the orthogonal code $\CC_k^\perp$ to the constraint code $\CC_k$ of Example 4, see \cite{GLS09}. \qed
\vspace{1ex}

Again, the point is that even though the HWGF of $\CC_i$ does not fully determine $\CC_i$, it does determine the HWGF of $\CC_i^\perp$, and \emph{vice versa}.

Ericson and Zinoviev \cite{EZ96} have generalized this method of obtaining further weight generating functions from complete weight generating functions, as follows.  A partition of a finite abelian group $G$ into disjoint subsets $\{G_j\}$ and of its character group $\hat{G}$ into disjoint subsets $\{\hat{G}_i\}$ is called a \emph{Fourier-invariant pair} if for all $i, j$ the transform of the indicator function $\Phi(G_j)$ of $G_j$, namely
$$
\hat{\Phi}(\hat{g}) = \sum_{g \in G} \Phi(G_j) \inner{\hat{g}}{g} = \sum_{g \in G_j} \inner{\hat{g}}{g},
$$
depends only on the subset $\hat{G}_i$ that contains $\hat{g}$, and similarly for the inverse transform, where $\inner{\hat{g}}{g} = \hat{g}(g)$.  For example, the Hamming partition pair defined by $G_0 = \{0\}, G_1 = G\setminus\{0\}$ and $\hat{G}_0 = \{0\}, \hat{G}_1 = \hat{G}\setminus\{0\}$ is Fourier-invariant.  For any Fourier-invariant partition pair, a MacWilliams identity may be obtained for the corresponding generalized weight generating function;  see \cite{EZ96} or \cite{F98}.  Zinoviev and Ericson \cite{ZE09} show that this concept is equivalent to that of an association scheme. 

\section{MacWilliams identities for terminated convolutional codes}

A principal use of MacWilliams identities is to compute the distance distribution of a linear code $\CC$ from that of its dual code $\CC^\perp$, where typically $\CC$ is high-rate and $\CC^\perp$ is low-rate (\ie $|\CC^\perp| < |\CC|$).  

For a convolutional code $\CC$, the most commonly studied distance distribution is its \emph{free (Hamming) distance spectrum}, namely, the distribution of (Hamming) weights of codewords in $\CC$ that start and end in the zero state without passing through an intermediate zero state.\footnote{We will assume in this section that the unique state sequence associated with the all-zero code sequence is the all-zero state sequence;  this assumption can always be satisfied by choosing a minimal encoder for $\CC$.}  Shearer and McEliece \cite{SM77} showed by example that the free distance spectrum of $\CC$ does not in general determine that of $\CC^\perp$, and therefore that there could be no MacWilliams identity for such distributions.

Recently, Bocharova, Hug, Johannesson and Kudryashov \cite{BHJK} have proved a MacWilliams identity  for truncations of a convolutional code $\CC$ and its orthogonal code $\CC^\perp$.  By letting the truncation length become large, they obtain an approximation to the free distance spectrum of $\CC$.

%\pagebreak

In this section, we derive similar results for weight distributions of codes obtained by various kinds of termination procedures, of which we regard tail-biting as the most elegant.  We argue that these alternative distributions are as useful for estimating code performance as the free distance spectrum.  These results effectively answer the original question posed by Shearer and McEliece \cite{SM77}, which we would state as follows:  is there a duality relationship that allows us to estimate the performance parameters of $\CC^\perp$ from those of $\CC$? 

\subsection{Terminated convolutional codes}

A general method for approximating the free distance spectrum of a linear convolutional code $\CC$ is to derive a series of block codes $\CC_N$ of length $N$ from $\CC$ by some sort of termination procedure, and then to study the distance distributions of $\CC_N$ as $N \to \infty$.  As we will see, for any of the termination methods below, the distance distribution of $\CC_N$, normalized by $N$, approaches the free distance spectrum of $\CC$ for $\dfree \le d < 2\dfree$, where $\dfree$ denotes the free distance of $\CC$ (the least weight of any nonzero code sequence).   However, we will argue that tail-biting is the nicest, particularly if we are also interested in the distance distribution of the orthogonal convolutional code $\CC^\perp$.  

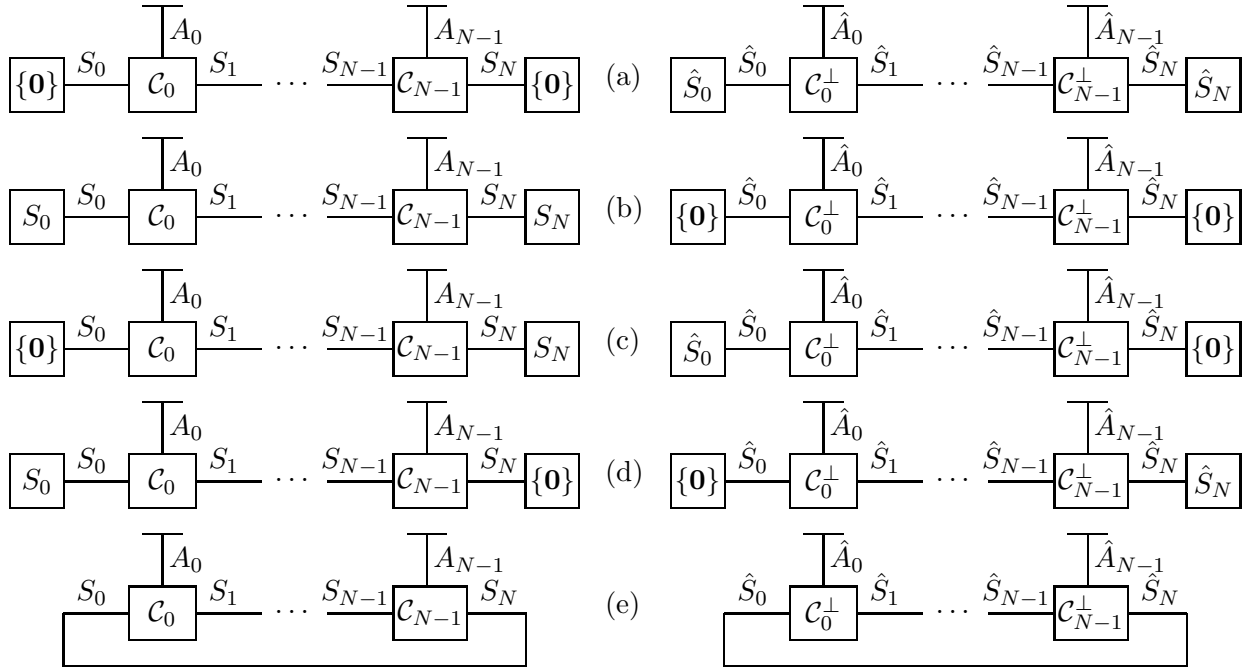
\begin{figure}[h]
\setlength{\unitlength}{5pt}
\centering
\begin{picture}(90,50)(-3, 1)
\put(0,5){\line(1,0){5}}
\put(0,5){\line(0,-1){4}}
\put(0,1){\line(1,0){35}}
\put(1,6){$S_0$}
\put(5,3){\framebox(5,4){$\CC_0$}}
\put(7.5,7){\line(0,1){4}}
\put(8,8.5){$A_0$}
\put(6,11){\line(1,0){3}}
\put(10,5){\line(1,0){5}}
\put(11,6){$S_{1}$}
\put(16,5){\ldots}
\put(20,5){\line(1,0){5}}
\put(19.5,6){$S_{N-1}$}
\put(25,3){\framebox(5.5,4){$\CC_{N-1}$}}
\put(27.5,7){\line(0,1){4}}
\put(28,8.5){$A_{N-1}$}
\put(26,11){\line(1,0){3}}
\put(30.5,5){\line(1,0){4.5}}
\put(31.5,6){$S_{N}$}
\put(35,5){\line(0,-1){4}}
\put(41,5){(e)}
\put(50,5){\line(0,-1){4}}
\put(50,1){\line(1,0){35}}
\put(50,5){\line(1,0){5}}
\put(51,6){$\hat{S}_0$}
\put(55,3){\framebox(5,4){$\CC_0^\perp$}}
\put(57.5,7){\line(0,1){4}}
\put(58,8.5){$\hat{A}_0$}
\put(56,11){\line(1,0){3}}
\put(60,5){\line(1,0){5}}
\put(61,6){$\hat{S}_{1}$}
\put(66,5){\ldots}
\put(70,5){\line(1,0){5}}
\put(69.5,6){$\hat{S}_{N-1}$}
\put(75,3){\framebox(5.5,4){$\CC_{N-1}^\perp$}}
\put(77.5,7){\line(0,1){4}}
\put(78,8.5){$\hat{A}_{N-1}$}
\put(76,11){\line(1,0){3}}
\put(80.5,5){\line(1,0){4.5}}
\put(81.5,6){$\hat{S}_{N}$}
\put(85,5){\line(0,-1){4}}
\put(46,13){\framebox(4,4){$\{\zerob\}$}}
\put(0,15){\line(1,0){5}}
\put(1,16){$S_0$}
\put(5,13){\framebox(5,4){$\CC_0$}}
\put(7.5,17){\line(0,1){4}}
\put(8,18.5){$A_0$}
\put(6,21){\line(1,0){3}}
\put(10,15){\line(1,0){5}}
\put(11,16){$S_{1}$}
\put(16,15){\ldots}
\put(20,15){\line(1,0){5}}
\put(19.5,16){$S_{N-1}$}
\put(25,13){\framebox(5.5,4){$\CC_{N-1}$}}
\put(27.5,17){\line(0,1){4}}
\put(28,18.5){$A_{N-1}$}
\put(26,21){\line(1,0){3}}
\put(30.5,15){\line(1,0){4.5}}
\put(31.5,16){$S_{N}$}
\put(35,13){\framebox(4,4){$\{\zerob\}$}}
\put(41,15){(d)}
\put(-4,13){\framebox(4,4){$S_0$}}
\put(50,15){\line(1,0){5}}
\put(51,16){$\hat{S}_0$}
\put(55,13){\framebox(5,4){$\CC_0^\perp$}}
\put(57.5,17){\line(0,1){4}}
\put(58,18.5){$\hat{A}_0$}
\put(56,21){\line(1,0){3}}
\put(60,15){\line(1,0){5}}
\put(61,16){$\hat{S}_{1}$}
\put(66,15){\ldots}
\put(70,15){\line(1,0){5}}
\put(69.5,16){$\hat{S}_{N-1}$}
\put(75,13){\framebox(5.5,4){$\CC_{N-1}^\perp$}}
\put(77.5,17){\line(0,1){4}}
\put(78,18.5){$\hat{A}_{N-1}$}
\put(76,21){\line(1,0){3}}
\put(80.5,15){\line(1,0){4.5}}
\put(81.5,16){$\hat{S}_{N}$}
\put(85,13){\framebox(4,4){$\hat{S}_N$}}
\put(-4,23){\framebox(4,4){$\{\zerob\}$}}
\put(0,25){\line(1,0){5}}
\put(1,26){$S_0$}
\put(5,23){\framebox(5,4){$\CC_0$}}
\put(7.5,27){\line(0,1){4}}
\put(8,28.5){$A_0$}
\put(6,31){\line(1,0){3}}
\put(10,25){\line(1,0){5}}
\put(11,26){$S_{1}$}
\put(16,25){\ldots}
\put(20,25){\line(1,0){5}}
\put(19.5,26){$S_{N-1}$}
\put(25,23){\framebox(5.5,4){$\CC_{N-1}$}}
\put(27.5,27){\line(0,1){4}}
\put(28,28.5){$A_{N-1}$}
\put(26,31){\line(1,0){3}}
\put(30.5,25){\line(1,0){4.5}}
\put(31.5,26){$S_{N}$}
\put(85,23){\framebox(4,4){$\{\zerob\}$}}
\put(41,25){(c)}
\put(46,23){\framebox(4,4){$\hat{S}_0$}}
\put(50,25){\line(1,0){5}}
\put(51,26){$\hat{S}_0$}
\put(55,23){\framebox(5,4){$\CC_0^\perp$}}
\put(57.5,27){\line(0,1){4}}
\put(58,28.5){$\hat{A}_0$}
\put(56,31){\line(1,0){3}}
\put(60,25){\line(1,0){5}}
\put(61,26){$\hat{S}_{1}$}
\put(66,25){\ldots}
\put(70,25){\line(1,0){5}}
\put(69.5,26){$\hat{S}_{N-1}$}
\put(75,23){\framebox(5.5,4){$\CC_{N-1}^\perp$}}
\put(77.5,27){\line(0,1){4}}
\put(78,28.5){$\hat{A}_{N-1}$}
\put(76,31){\line(1,0){3}}
\put(80.5,25){\line(1,0){4.5}}
\put(81.5,26){$\hat{S}_{N}$}
\put(35,23){\framebox(4,4){$S_N$}}
\put(46,33){\framebox(4,4){$\{\zerob\}$}}
\put(0,35){\line(1,0){5}}
\put(1,36){$S_0$}
\put(5,33){\framebox(5,4){$\CC_0$}}
\put(7.5,37){\line(0,1){4}}
\put(8,38.5){$A_0$}
\put(6,41){\line(1,0){3}}
\put(10,35){\line(1,0){5}}
\put(11,36){$S_{1}$}
\put(16,35){\ldots}
\put(20,35){\line(1,0){5}}
\put(19.5,36){$S_{N-1}$}
\put(25,33){\framebox(5.5,4){$\CC_{N-1}$}}
\put(27.5,37){\line(0,1){4}}
\put(28,38.5){$A_{N-1}$}
\put(26,41){\line(1,0){3}}
\put(30.5,35){\line(1,0){4.5}}
\put(31.5,36){$S_{N}$}
\put(85,33){\framebox(4,4){$\{\zerob\}$}}
\put(41,35){(b)}
\put(-4,33){\framebox(4,4){$S_0$}}
\put(50,35){\line(1,0){5}}
\put(51,36){$\hat{S}_0$}
\put(55,33){\framebox(5,4){$\CC_0^\perp$}}
\put(57.5,37){\line(0,1){4}}
\put(58,38.5){$\hat{A}_0$}
\put(56,41){\line(1,0){3}}
\put(60,35){\line(1,0){5}}
\put(61,36){$\hat{S}_{1}$}
\put(66,35){\ldots}
\put(70,35){\line(1,0){5}}
\put(69.5,36){$\hat{S}_{N-1}$}
\put(75,33){\framebox(5.5,4){$\CC_{N-1}^\perp$}}
\put(77.5,37){\line(0,1){4}}
\put(78,38.5){$\hat{A}_{N-1}$}
\put(76,41){\line(1,0){3}}
\put(80.5,35){\line(1,0){4.5}}
\put(81.5,36){$\hat{S}_{N}$}
\put(35,33){\framebox(4,4){$S_N$}}
\put(-4,43){\framebox(4,4){$\{\zerob\}$}}
\put(0,45){\line(1,0){5}}
\put(1,46){$S_0$}
\put(5,43){\framebox(5,4){$\CC_0$}}
\put(7.5,47){\line(0,1){4}}
\put(8,48.5){$A_0$}
\put(6,51){\line(1,0){3}}
\put(10,45){\line(1,0){5}}
\put(11,46){$S_{1}$}
\put(16,45){\ldots}
\put(20,45){\line(1,0){5}}
\put(19.5,46){$S_{N-1}$}
\put(25,43){\framebox(5.5,4){$\CC_{N-1}$}}
\put(27.5,47){\line(0,1){4}}
\put(28,48.5){$A_{N-1}$}
\put(26,51){\line(1,0){3}}
\put(30.5,45){\line(1,0){4.5}}
\put(31.5,46){$S_{N}$}
\put(35,43){\framebox(4,4){$\{\zerob\}$}}
\put(41,45){(a)}
\put(46,43){\framebox(4,4){$\hat{S}_0$}}
\put(50,45){\line(1,0){5}}
\put(51,46){$\hat{S}_0$}
\put(55,43){\framebox(5,4){$\CC_0^\perp$}}
\put(57.5,47){\line(0,1){4}}
\put(58,48.5){$\hat{A}_0$}
\put(56,51){\line(1,0){3}}
\put(60,45){\line(1,0){5}}
\put(61,46){$\hat{S}_{1}$}
\put(66,45){\ldots}
\put(70,45){\line(1,0){5}}
\put(69.5,46){$\hat{S}_{N-1}$}
\put(75,43){\framebox(5.5,4){$\CC_{N-1}^\perp$}}
\put(77.5,47){\line(0,1){4}}
\put(78,48.5){$\hat{A}_{N-1}$}
\put(76,51){\line(1,0){3}}
\put(80.5,45){\line(1,0){4.5}}
\put(81.5,46){$\hat{S}_{N}$}
\put(85,43){\framebox(4,4){$\hat{S}_N$}}
\end{picture}
\caption{Terminated convolutional codes and their duals (over a field of characteristic 2).}
\label{FigTerm}
\end{figure}

The left side of Figure \ref{FigTerm} shows normal graphs of five block codes $\CC_N$ obtained from a convolutional code $\CC$ by five kinds of termination procedures: 
\begin{itemize}
\item[(a)] the subcode $\CC_{[0,N)}$;  
\item[(b)] the projection $\CC_{|[0,N)}$;  
\item[(c)] the truncated code $\CC_{\lhd [0,N)}$; 
\item[(d)]  the reverse-truncated code $\CC_{\rhd [0,N)}$;  and 
\item[(e)] the tail-biting code $\CC_{|| [0,N)}$.  
\end{itemize}
In each case the central part of the graph consists of $N$ consecutive trellis sections of $\CC$, and the block code symbols are the corresponding convolutional code symbols $(a_0, \ldots, a_{n-1})$.  

To obtain the subcode $\CC_{[0,N)}$,  the starting and ending state variables are constrained to be zero:  $s_0 = s_N = 0$.  For the projection $\CC_{|[0,N)}$, the starting and ending states may be any arbitrary pair $s_0 \in S_0, s_N \in S_N$.  For the truncated code $\CC_{\lhd [0,N)}$, we constrain $s_0 = 0$, but let $s_N$ be arbitrary;  for the reverse-truncated code,  the reverse constraints are imposed.  Finally, for the tail-biting code $\CC_{|| [0,N)}$, we impose the constraint $s_N = s_0$.

The right side of Figure \ref{FigTerm} shows the orthogonal codes to all of these codes, obtained simply by use of the normal graph duality theorem, along with the observation that the orthogonal code to a trivial code $\{\zerob\}$ is a universe code, \eg $\hat{S}_0$, and \emph{vice versa}.  (For simplicity, we have assumed that the characteristic of $\F$ is 2, so that we do not need to show sign inverters.)  

We observe that the orthogonal code to a subcode $\CC_{[0,N)}$ of $\CC$ is the corresponding projection $(\CC^\perp)_{|[0,N)}$ of $\CC^\perp$, and \emph{vice versa}, as is well known;  the orthogonal code to a truncated code $\CC_{\lhd [0,N)}$ of $\CC$ is the corresponding reverse-truncated code $(\CC^\perp)_{\rhd[0,N)}$ of  $\CC^\perp$, and \emph{vice versa} \cite{BHJK};  and the orthogonal code to a tail-biting code $\CC_{||[0,N)}$ of $\CC$ is the corresponding tail-biting code $(\CC^\perp)_{||[0,N)}$ of $\CC^\perp$ \cite{F01}. 

Since $\CC_{[0,N)}$ and $(\CC^\perp)_{|[0,N)}$, or $\CC_{\lhd[0,N)}$ and $(\CC^\perp)_{\rhd[0,N)}$, or  $\CC_{||[0,N)}$ and $(\CC^\perp)_{||[0,N)}$ are orthogonal block codes, there is a MacWilliams identity between their respective weight generating functions. 

%\pagebreak
We now give examples that will illustrate  these various kinds of terminated codes, and the following general properties:
\begin{itemize}
\item The rate of a subcode $\CC_{[0,N)}$ is less than that of $\CC$, the rate of a projection is higher than that of $\CC$, and the rate of a truncated or a tail-biting code is the same as that of $\CC$.
\item The minimum distance of a subcode $\CC_{[0,N)}$ is (at least) the same as the minimum free distance $\dfree$ of $\CC$.  As is well known, a tail-biting code has the same property, provided that $N$ is large enough.  However, the  other kinds of terminated codes in general have low-weight codewords.
\end{itemize}

\vspace{1ex}
\noindent
\textbf{Example 3} (cont.).  We now consider various methods of terminating the rate-$\half$   binary convolutional code $\CC$ of Example 3 with a block length of $N = 4$.  

The subcode $\CC_{[0,4)}$ is the $(8,2)$ binary linear block code generated by the two generators
$$
\begin{array}{cccc}
11 & 01 & 11 & 00 \\
00 & 11 & 01 & 11 \\
\end{array}
$$
The minimum distance of this block code is the same as the minimum free distance $\dfree = 5$ of $\CC$, although its rate is lower.

%\pagebreak
The orthogonal code to the subcode $\CC_{[0,4)}$ is the projection $(\CC^\perp)_{|[0,4)}$ of the orthogonal convolutional code $\CC^\perp$, which is the  $(8,6)$ binary linear block code generated by the six generators
$$
\begin{array}{cccc}
11 & 00 & 00 & 00 \\
10 & 11 & 00 & 00 \\
11 & 10 & 11 & 00 \\
00 & 11 & 10 & 11 \\
00 & 00 & 11 & 10 \\
00 & 00 & 00 & 11 \\
\end{array}
$$
The minimum distance of this block code is 2, less than the minimum free distance of $\CC^\perp$, although its rate is higher.

The truncated code $\CC_{\lhd[0,4)}$ is the $(8,4)$ binary linear block code generated by
$$
\begin{array}{cccc}
11 & 01 & 11 & 00 \\
00 & 11 & 01 & 11 \\
00 & 00 & 11 & 01 \\
00 & 00 & 00 & 11 \\
\end{array}
$$
The minimum distance of this block code is 2, but its rate is the same as that of $\CC$.  Its orthogonal code  $(\CC^\perp)_{\rhd[0,4)}$ is the $(8,4)$ binary linear block code generated by
$$
\begin{array}{cccc}
11 & 00 & 00 & 00 \\
10 & 11 & 00 & 00 \\
11 & 10 & 11 & 00 \\
00 & 11 & 10 & 11 \\
\end{array}
$$
which has the same parameters.

The tail-biting terminated code $\CC_{||[0,4)}$ is the $(8,4)$ binary linear block code generated by
$$
\begin{array}{cccc}
11 & 01 & 11 & 00 \\
00 & 11 & 01 & 11 \\
11 & 00 & 11 & 01 \\
01 & 11 & 00 & 11  
\end{array}
$$
whereas the orthogonal tail-biting terminated code $(\CC^\perp)_{||[0,N)}$ is the $(8,4)$ binary linear block code generated by the four generators
$$
\begin{array}{cccc}
11 & 10 & 11 & 00 \\
00 & 11 & 10 & 11 \\
11 & 00 & 11 & 10 \\
10 & 11 & 00 & 11  
\end{array}
$$
Both of these codes have a minimum distance of only 2 (\eg for paths such as 01 00 01 00 from state 10 to state 10 in $\CC_{||[0,4)}$).  However, for $N \ge 10$, it turns out that the minimum distance of both tail-biting terminated codes is 5, the same as the minimum free distance of $\CC$ or $\CC^\perp$. \qed \vspace{1ex}

\subsection{Distance distributions of convolutional codes and terminated codes}

In this subsection, we consider how the free distance spectrum of a linear time-invariant convolutional code $\CC$ may be derived from the weight distribution of any of these terminated codes of length $N$ as $N \to \infty$.   Then, in the next subsection, we show how the weight generating functions of any of these terminated codes may be determined from the weight adjacency matrix of the convolutional code.  Again, the most elegant relationships are obtained for tail-biting terminated codes.

We continue to assume that the unique state sequence associated with the infinite all-zero code sequence is the all-zero state sequence.  Consequently, the lowest-weight words of a terminated code as $N \to \infty$ must be those that pass through the zero state almost all of the time.  These code sequences are as follows, for the various termination methods we have considered:
\begin{itemize}
\item If we terminate to the subcode $\CC_{[0,N)}$, then code sequences start and end in the zero state, and the lowest-weight sequences correspond to the lowest-weight sequences in the free distance spectrum.  If the minimum free distance is $\dfree$, then for $\dfree \le d < 2\dfree$ there will be approximately $N \times N_d$ sequences in the terminated code of weight $d$, where $N_d$ is the number of code sequences of weight $d$ in the free distance spectrum of $\CC$.  Thus, for $\dfree \le d < 2\dfree$, the weight distribution per unit time of $\CC$ is the limit of the weight distribution of $\CC_{[0,N)}$ normalized by (divided by) $N$ as $N \to \infty$.  For $d \ge 2\dfree$, there will be overcounting--- \eg two sequences of weight $\dfree$ may be counted as one of weight $2\dfree$--- but we will argue below that such overcounting should not affect estimates of code performance.
\item If we terminate to the projection $\CC_{|[0,N)}$, then code sequences can start and end in any state, and there will be low-weight sequences starting with a low-weight state transition $s \to 0$, remaining in state 0 for nearly $N$ time units, and then ending with a low-weight transition $0 \to s'$, where $s$ and $s'$ are not both 0.  Thus the minimum distance of  $\CC_{|[0,N)}$ will be less than $\dfree$ for all $N$.  However, the number of such low-weight sequences remains constant, so after normalization we will eventually see the same normalized weight distribution as for $\CC_{[0,N)}$.
\item If we terminate to the truncated code $\CC_{\lhd[0,N)}$, then by the same argument we will eventually see the correct normalized weight distribution.  In this case, for a code sequence that starts in the zero state, remains there for nearly $N$ time units, and then ends with a low-weight transition $0 \to s$, the total weight is only that of the low-weight transition  $0 \to s$.    However, again the number of such low-weight sequences remains constant, so after normalization we will eventually see the correct normalized weight distribution.
\item If we terminate to the tail-biting code $\CC_{||[0,N)}$, then by the same argument we will eventually see the correct normalized weight distribution.  Note however that in this case the total weight of a code sequence starting with a low-weight transition $s \to 0$, remaining in the zero state for nearly $N$ time units, and then ending with a low-weight transition $0 \to s$, must be at least $\dfree$, since the ending sequence (corresponding to the state transition $0 \to s$) followed by the starting sequence (corresponding to $s \to 0$) must be a cyclic shift of a code sequence of $\CC$.  Thus the minimum distance of  $\CC_{||[0,N)}$ must equal $\dfree$ for large enough $N$.
\end{itemize} 

We conclude that as $N \to \infty$ the normalized weight distribution of any of these terminated codes approaches the free distance spectrum of $\CC$ for $\dfree \le d < 2\dfree$.  However, only the tail-biting code has the same rate as $\CC$ and the same minimum distance $\dfree$ (for large enough $N$).

We now argue that the normalized weight distribution of any of these terminated codes $\CC_N$ must yield the same estimate of code performance over $N$ time units as the free distance spectrum of $\CC$, if these estimates are accurate.  The probability of error event $P(\E)$ of $\CC$ per unit time may be estimated using the free distance spectrum.  The probability of any error in $N$ time units is then estimated as $NP(\E)$.  If this is a good estimate (implying $N < 1/P(\E)$), then the probability of two or more error events in $N$ time units must be negligible.  But the probability of any error in decoding $\CC$ over $N$ time units is essentially the same as the probability of block decoding error in decoding $\CC_N$, which may be estimated by the weight distribution of $N$, which counts codewords that include two or more error events.  If the probability of two or more error events in $N$ time units is negligible, then an estimate based on the weight distribution of $\CC_N$ must approximately agree with an estimate based on the free distance spectrum of $\CC$.

\subsection{Free distance spectra for convolutional codes from terminated codes}

We now show how weight generating functions for terminations of a linear time-invariant convolutional code $\CC$ may be derived from the weight adjacency matrix of the constraint code $\CC_k$ that specifies $\CC$.  This will allow us to state MacWilliams identities for terminated convolutional codes, and to estimate code performance.

%As a subcode of $\CC$, $\CC_{[0,N)}$ has the same minimum distance as the minimum free distance $\dfree$ of $\CC$.   Similarly, for large enough $N$, the low-weight codewords of the tail-biting code $\CC_{||[0,N)}$ must be cyclic shifts of code sequences of $\CC$ (under our assumption that the only infinite-length code sequence associated with the all-zero state sequence is the all-zero sequence), so for large enough $N$ the minimum distance of $\CC_{||[0,N)}$ is $\dfree$.  However, the other terminated codes contain codewords that come from arbitrary non-zero states to the zero state or \emph{vice versa}, and so in general contain low-weight  codewords, no matter how large $N$ becomes.

We compute the Hamming weight distributions of these terminated codes as follows.  Let $\Lambda_{[0,N)}(x)$ be the Hamming weight adjacency matrix of $\CC$ over the interval $[0,N)$, whose elements are indexed by $S_0 \times S_N$.  As we have seen in Example 2, if $\CC$ is time-invariant and $\Lambda(x)$ is the HWAM of each constraint code $\CC_k$, then $\Lambda_{[0,N)}(x)$ is simply equal to $\Lambda^N(x)$.  

From their definitions, we see that the HWGFs of various terminated codes of $\CC$ can be read as follows from the HWAM $\Lambda^N(x)$:
\begin{itemize}
\item[(a)] The HWGF of the subcode $\CC_{[0,N)}$ is the $(0,0)$ element of $\Lambda^N(x)$.
\item[(b)] The HWGF of the projection $\CC_{|[0,N)}$ is the sum of all elements of $\Lambda^N(x)$.
\item[(c)]  The HWGF of the truncated code $\CC_{\lhd[0,N)}$ is the sum of all elements in the first row of $\Lambda^N(x)$.
\item[(d)]  The HWGF of the reverse-truncated code $\CC_{\rhd[0,N)}$ is the sum of all elements in the first column of $\Lambda^N(x)$.
\item[(e)] The HWGF of the tail-biting code $\CC_{||[0,N)}$ is the sum of all diagonal elements of $\Lambda^N(x)$;  \ie its trace $\Tr(\Lambda^N(x))$ (see Example 5).  
\end{itemize} 

\vspace{1ex}
\noindent
\textbf{Example 3} (cont.).  For the rate-1/2 binary convolutional code $\CC$ of Example 3, the HWAM of a section consisting of $N = 2$ time units of our example code $\CC$ is thus
$$
\Lambda^2(x) \quad  = \quad
\left[
\begin{array}{cccc}
1 & x^2 & x^3 & x^3 \\
x^3 & x & x^2 & x^2 \\
x^2 & x^4 & x & x \\
x^3 & x & x^2 & x^2 
\end{array}
\right].
$$
This shows that there is exactly one path from each state in $S_k$ to each state in $S_{k+2}$, and that the minimum Hamming weight of any of these paths (other than the zero path) is 1.

For a section consisting of $N = 4$ time units of this code, the HWAM is
$$
\Lambda^4(x)  =
\left[
\begin{array}{cccc}
1 + 2x^5 + x^6 & x^2 + x^3 + x^4 + x^7 & x^3 + 2x^4 + x^5 & x^3 + 2x^4 + x^5  \\
x^3 + 2 x^4 + x^5 & x^2 + x^3 +  x^5 + x^6 & 2x^3 + x^4 + x^6 & 2x^3 + x^4 + x^6 \\
x^2 + x^3 + x^4 + x^7 & x^2 +  x^4 + 2x^5 & x^2 + x^3 + x^5 + x^6 & x^2 + x^3 + x^5 + x^6 \\
x^3 + 2 x^4 + x^5 & x^2 + x^3 +  x^5 + x^6 &2x^3 + x^4 + x^6 & 2x^3 + x^4 + x^6
\end{array}
\right].
$$
This shows that there are four paths from each state in $S_k$ to each state in $S_{k+4}$, and that the minimum nonzero Hamming weight of any of these paths is 2. 

The Hamming weight generating function of the tail-biting termination $\CC_{||[0,4)}$ of length 4 is the trace of $\Lambda^4(x)$, namely $1 + 2x^2 + 4x^3 + x^4 + 4x^5 + 4x^6$.  Since $\hat{\Lambda}^4(x)$ happens to be the transpose of $\Lambda^4(x)$, the orthogonal tail-biting terminated code $(\CC^\perp)_{||[0,4)}$ has the same Hamming weight generating function.  It is easy to check that the Hamming weight generating function of this code is indeed invariant under the MacWilliams transform. \qed \vspace{1ex}

Using tail-biting terminated codes, and normalizing the weight distribution by dividing by $N$, we have that the generating function of the normalized Hamming weight distribution of $\CC$ is
$$
g_{\CC}(x) = \lim_{N \to \infty} \frac{1}{N} \Tr(\Lambda^N(x)).
$$
Moreover, there is a MacWilliams identity between $g_{\CC}(x)$ and $g_{\CC^\perp}(x)$.  The performance of $\CC$ may be estimated from $g_{\CC}(x)$, and that of $\CC^\perp$ from $g_{\CC^\perp}(x)$.
(Similar observations are made in \cite{BHJK}, using truncated codes.)

It appears that the behavior of $g_{\CC}(x)$ might be analyzed by using an extension of Perron-Frobenius theory to generating function matrices, as in  \cite{FKMT}; however, we have not attempted such an analysis.

\vspace{1ex}
\noindent
\textbf{Example 1} (cont.).  For a section consisting of $N = 16$ time units of the rate-1/2 binary convolutional code $\CC$ of Example 1, the HWAM $\Lambda^{16}(x)$ (modulo $x^8$) is 
$$\scriptsize
\begin{array}{cccc}
1 + 14x^5 + 25x^6 + 44x^7 & x^2 + x^3 + 2x^4 +4x^5 +8x^6 + 29x^7 & x^3 + 2 x^4 + 4x^5 + 8x^6 + 16x^7 & x^3 + 2 x^4 + 4x^5 + 8x^6 + 16x^7  \\
x^3 + 2 x^4 + 4x^5 + 8x^6 + 16x^7 & x^5 + 3x^6 + 8x^7 & x^6 + 4x^7 & x^6 + 4x^7 \\
x^2 + x^3 + 2x^4 +4x^5 +8x^6 + 29x^7 & x^4 + 2x^5 +5x^6 +12x^7 & x^5 + 3x^6 + 8x^7 & x^5 + 3x^6 + 8x^7 \\
x^3 + 2 x^4 + 4x^5 + 8x^6 + 16x^7 & x^5 + 3x^6 + 8x^7 & x^6 + 4x^7 & x^6 + 4x^7
\end{array}
$$
Notice that 
$$\Tr(\Lambda^{16}(x)) = 1 + 16x^5 + 32x^6 + 64x^7 + \cdots,$$
 so that normalizing the distribution by dividing the higher-order coefficients by $N = 16$ already gives the precise free distance spectrum of $\CC$ for $d < 8$, namely $x^5 + 2x^6 + 4x^7 + \cdots$. Thus the convergence to the limiting generating function $g_{\CC}(x)$ is rapid and exact.  This property of tail-biting codes is not shared by other kinds of terminations. \qed \vspace{1ex}

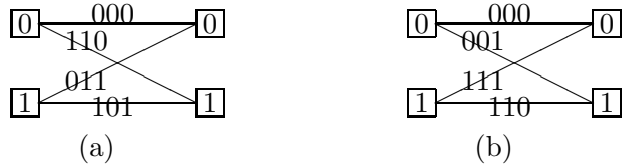
\begin{figure}[h]
\setlength{\unitlength}{5pt}
\centering
\begin{picture}(42,8)(0, -1)
\put(0,0){\framebox(2,2){$1$}}
\put(0,6){\framebox(2,2){$0$}}
\put(14,0){\framebox(2,2){$1$}}
\put(14,6){\framebox(2,2){$0$}}
\put(2,1){\line(1,0){12}}
\put(2,1){\line(2,1){12}}
\put(2,7){\line(2,-1){12}}
\put(2,7){\line(1,0){12}}
\put(6,0){$101$}
\put(4,5){$110$}
\put(6,7){$000$}
\put(4,2){$011$}
\put(5,-3){(a)}

\put(30,0){\framebox(2,2){$1$}}
\put(30,6){\framebox(2,2){$0$}}
\put(44,0){\framebox(2,2){$1$}}
\put(44,6){\framebox(2,2){$0$}}
\put(32,1){\line(1,0){12}}
\put(32,1){\line(2,1){12}}
\put(32,7){\line(2,-1){12}}
\put(32,7){\line(1,0){12}}
\put(36,0){$110$}
\put(34,5){$001$}
\put(36,7){$000$}
\put(34,2){$111$}
\put(35,-3){(b)}
\end{picture}

\caption{Trellis sections of (a) rate-1/3 2-state binary convolutional code $\CC_1$; (b) similar code $\CC_2$.}
\label{FigSM}
\end{figure}

\vspace{1ex}\pagebreak
\noindent
\textbf{Example 6} (\cf \cite{SM77, BHJK}).  The two codes proposed by Shearer and McEliece \cite{SM77} for their counterexample provide an excellent final example.  The first code is a rate-1/3 binary linear time-invariant convolutional code $\CC_1$ generated by the degree-1 generators $(1, 1 + D, D)$, \ie $\CC_1$ is generated by a minimal encoder with impulse response is $(110, 011, 000, \ldots)$, whose trellis section is shown in Figure \ref{FigSM}(a).  The HWAM of this encoder is
$$
\Lambda_1(x) \quad  = \quad
\left[
\begin{array}{cc}
1 & x^2 \\
x^2 & x^2 \\
\end{array}
\right].
$$

The second code is a rate-1/3 binary linear time-invariant convolutional code $\CC_2$ generated by the degree-1 generators $(D, D, 1 + D)$, \ie $\CC_2$ is generated by a minimal encoder whose impulse response is $(001, 111, 000, \ldots)$, whose trellis section is shown in Figure \ref{FigSM}(b).  The HWAM of this encoder is
$$
\Lambda_2(x) \quad  = \quad
\left[
\begin{array}{cc}
1 & x \\
x^3 & x^2 \\
\end{array}
\right].
$$

Since the weights of the $0 \to 0$ and $1 \to 1$ transitions are the same for $\CC_1$ and $\CC_2$, and since the sums of the weights of the $0 \to 1$ and $1 \to 0$ transitions are the same, it is evident that the weight distributions of the subcodes $(\CC_1)_{[0,N)}$ and $(\CC_2)_{[0,N)}$ are the same for all $N$, and that the free distance spectra of $\CC_1$ and $\CC_2$ are also the same.  For the same reason, the weight distributions of the tail-biting codes $(\CC_1)_{||[0,N)}$ and $(\CC_2)_{||[0,N)}$ are the same for all $N$. 

However, the weight distributions of the projections $(\CC_1)_{|[0,N)}$ and $(\CC_2)_{|[0,N)}$ are not the same even for $N = 1$. It follows that the weight distributions of the subcodes $(\CC_1^\perp)_{[0,N)}$ and $(\CC_2^\perp)_{[0,N)}$ of their orthogonal rate-2/3 codes $\CC_1^\perp$ and $\CC_2^\perp$ are not the same, and therefore that their free distance spectra are not the same;  this was the point of Shearer and McEliece \cite{SM77}.

On the other hand, since the weight distributions of the tail-biting codes $(\CC_1)_{||[0,N)}$ and $(\CC_2)_{||[0,N)}$ are the same for all $N$, it follows that \emph{the weight distributions of the tail-biting codes $(\CC_1^\perp)_{||[0,N)}$ and $(\CC_2^\perp)_{||[0,N)}$ are the same for all $N$.}

  Since the performance of $\CC_1^\perp$ and $\CC_2^\perp$ may be estimated from these weight distributions as $N \to \infty$, it follows that the performance of $\CC_1^\perp$ and $\CC_2^\perp$ is effectively the same, despite the difference in their free distance spectra.\footnote{Another way of reaching the same (or a stronger) conclusion is to observe that $\CC_1$ and $\CC_2$ are equivalent under a simple finite-memory permutation.  Therefore $\CC_1^\perp$ and $\CC_2^\perp$ must be equivalent under the same permutation, and thus must have precisely the same performance on a memoryless channel with maximum likelihood decoding.} \qed \vspace{1ex}

In summary, similarly to \cite{BHJK}, we have shown that there is a MacWilliams identity between the generating functions of the weight distributions per unit time of a linear convolutional code $\CC$ and its orthogonal code $\CC^\perp$ (as calculated from their tail-biting terminations), and that these distributions are as useful as their free distance spectra for estimating code performance.  These results effectively answer the question posed by Shearer and McEliece \cite{SM77}.

\section*{Acknowledgments}  
For discussions that led to the conceptual framework of partition functions of normal factor graphs, I am deeply grateful to Pascal Vontobel and Yongyi Mao. 
I also thank Heide Gluesing-Luerssen, Rolf Johannesson, Pascal Vontobel and the reviewers for many helpful comments on earlier versions of this paper.
%\pagebreak

\section*{Author biography}

\textbf{G. David Forney, Jr.} received the B.S.E. degree in electrical engineering from Princeton University, Princeton, NJ, in 1961, and the M.S. and Sc.D. degrees in electrical engineering from the Massachusetts Institute of Technology (M.I.T.), Cambridge, MA, in 1963 and 1965, respectively.

	From 1965-99 he was with the Codex Corporation, which was acquired by Motorola, Inc. in 1977, and its successor, the Motorola Information Systems Group, Mansfield, MA.  Since 1996, he has been an Adjunct Professor at M.I.T.

	Dr. Forney was Editor of the IEEE Transactions on Information Theory from 1970 to 1973.  He has been a member of the Board of Governors of the IEEE Information Theory Society during 1970-76, 1986-94, and 2004-10, and was President in 1992 and 2008.  He has been awarded the 1970 IEEE Information Theory Group Prize Paper Award, the 1972 IEEE Browder J. Thompson Memorial Prize Paper Award, the 1990 and 2009 IEEE Donald G. Fink Prize Paper Awards, the 1992 IEEE Edison Medal, the 1995 IEEE Information Theory Society Claude E. Shannon Award, the 1996 Christopher Columbus International Communications Award, and the 1997 Marconi International Fellowship.  In 1998 he received an IT Golden Jubilee Award for Technological Innovation, and two IT Golden Jubilee Paper Awards.  He received an honorary doctorate from EPFL, Lausanne, Switzerland in 2007.  He was elected a Fellow of the IEEE in 1973, a member of the National Academy of Engineering (U.S.A.) in 1983, a Fellow of the American Association for the Advancement of Science in 1993, an honorary member of the Popov Society (Russia) in 1994, a Fellow of the American Academy of Arts and Sciences in 1998, and a member of the National Academy of Sciences (U.S.A.) in 2003.

\end{document}